\documentstyle[prb,preprint,aps]{revtex}
\font\ee=msbm10 scaled \magstep1
\font\ee=msbm10 scaled \magstep1

\begin{document}
\title
{\bf Operational approach in the weak-field measurement   
of polarization fluctuations} 
\author{T. Hakio\u{g}lu} 
\address{Physics Department, Bilkent University, 
06533-Ankara, Turkey} 
\date{\today}
\maketitle
\begin{abstract}
The operational approach to the measurement of phase studied by 
Noh, Fouger\`es and Mandel is 
applied to the measurement of the state of polarization of fully 
polarized light. Operational counterparts of the quantum Stokes parameters 
are introduced and their fluctuations are examined. 
It is shown that, if the polarized field is 
weak, the measured fluctuations are influenced not only by the quantum 
properties of the source field but also that of the measurement. This    
character is reflected on the measured probability distributions of the 
parameters of polarization, which are also investigated independently   
for the fully polarized coherent states and the Fock states as the initial 
field strength is varied. Finally, connection between the operational approach 
to the measurement of polarization and the $su(2)$ interferometry is 
examined. 
\end{abstract}
~~~~~~~PACS numbers:\ 42.50, 42.50.B, 42.50.D, 85.60.G
\vfill
\section{Introduction}
The idea of the operational approach as an experimental technique based 
on photon counting in the measurement of quantum phase 
fluctuations has been suggested in the late 80s by Barnett and Pegg\cite{1} 
in the context of a measured phase operator using certain homodyne 
experiments and more recently 
was formulated in detail by Noh, Fouger\'es and Mandel (NFM)\cite{2}. The 
operational phase measurement is based on using N-port quantum homodyne 
detectors of which the analogy with classical homodyne approach is based on  
the purpose of extracting information about the 
phase between two initial fields 
by performing a complete set of photocount measurements 
between the components of the field. This procedure of obtaining the phase 
information 
between two fields depends on the particular experimental scheme through 
its classical analogy of relating the relative photocount measurements to 
certain cosine and sine functions of the relative phase. Since, through this  
suggested analogy, different quantum measurement schemes 
would correspond to different classical ones, the information extracted for 
the relative phase is expected to be different 
for different experimental schemes. Indeed this point has been 
demonstrated in the formulation of the operational phase measurement by NFM 
by starting with two different classical and quantum measurement schemes 
where one measurement used two-port homodyne detection whereas the second one 
used four-port homodyne detection\cite{2,3}. The two port measurement yields 
either the cosine or the sine information about the phase failing to 
give the full phase information. In the four port scheme the simultaneous 
measurements were made possible by well-defined trigonometric operators of 
the relative phase where the full information on the phase and its 
fluctuations can be extracted. On the other hand a comparison of NFM's  
operational approach with the operational approach introduced by Vogel  
and Schleich\cite{4} has been compared by Lynch\cite{5} who found agreement 
between the two operational schemes.  

Another particularly important part of this scheme dependence 
manifests itself in the weak-field measurements in which the quantized 
nature of light as well as of the detectors becomes crucial when  
the homodyne detectors have a relatively high probability of registering a 
few or null photocounts within the measurement time interval $T$. This being 
the case for a single 
set of measurements, one considers an ensemble of repeated measurements 
under the same initial conditions. There, each repeated measurement would 
have generally different but equally acceptable configurations of detected 
photons and one has to make a 
distinction between the outcome of a single measurement from the average 
outcome of a collection of such repeated measurements under otherwise the 
same conditions. Despite the fact that the experimental verification of the 
operational NFM approach was successfully made by the same group\cite{6}, the 
appearance of the discrete outcomes in the phase measurements in their scheme 
was subject to long and heavy discussions\cite{7,8,9,10}. In this work we 
suggest another application of their approach to the operational 
measurement of the state of polarization of a fully polarized source. 
In an earlier publication,\cite{11} we investigated  
a particular extension of NFM's operational 
approach to the measurement of the Stokes parameters 
of a fully polarized weak coherent light. In this work, 
we will extend this formalism introduced in [11] to a more general   
framework by including the calculations for the measured probability 
distributions of the polarization fluctuations and also examine the case 
throughly when the initial field is a fully polarized Fock state.  

We start with a brief outline  
of the operational approach to the measurement of  
polarization fluctuations when the polarized field is given in a classical 
as well as a quantum state. In Sec.\,II we present the 
general formalism of calculating the polarization fluctuations and their 
corresponding 
probability distributions. Sections II.A and II.B are devoted to the 
specific calculations corresponding to two different fully polarized 
initial quantum states of the field as coherent and Fock states 
respectively. Section II.C is devoted to the connections between the 
operational approach and the $su(2)$ interferometry. 

Classically, the state of polarization of a fully polarized monochromatic  
field, $E_{i}=\epsilon_{i}\,\cos{(\omega\,t+\delta_{i})}$, 
where $i=1,2$ are the polarization indices of two preselected 
orthogonal polarization eigenmodes, can be manifestly described by 
four Stokes parameters $s_{m}~~(m=0,1,2,3)$ as\cite{12,13,14,15} 
\begin{equation}
\begin{array}{rl}
s_{0}=&\frac{1}{2}\,(\langle E_{1}^{2}\rangle+(\langle E_{2}^{2}\rangle)~,\\
\\
s_{1}=&\frac{1}{2}\,(\langle E_{1}^{2}\rangle-(\langle E_{2}^{2}\rangle)~,\\
\\
s_{2}=&(\langle E_{1}^{2}\rangle\,\langle E_{2}^{2}\rangle)^{1/2}\,
\cos{\phi}~, \\
\\
s_{3}=&(\langle E_{1}^{2}\rangle\,\langle E_{2}^{2}\rangle)^{1/2}\,
\sin{\phi}  
\end{array}
\label{I.1}
\end{equation}
where $\phi=\delta_{2}-\delta_{1}$ is the optical (temporal) phase and 
$I_{i}=\langle E_{i}^{2}\rangle$ is the intensity of the corresponding  
$i$th component (~$i=1,2$). We now describe an experimental setup  
based on a set of photocount measurements for the purpose of investigating 
the fluctuations in the measurement of the classical Stokes parameters     
in Eqs\,(\ref{I.1}) and their corresponding quantum counterparts.  

\subsection{The Classical Measurement Scheme}
Within the operational approach, it is possible to measure all classical 
Stokes parameters in terms of the various components of the 
intensity. The experimental scheme is shown in Fig.\,(1). The initial 
field enters the setup through the $\% 50-\% 50$ beam splitter $BS_{1}$. 
One of the output beams of $BS_{1}$ is sent to a polarizing beam 
splitter $PBS_{1}$ which defines a reference frame $1, 2$ for the      
relative angular orientation of all other polarizing beam splitters. 
The other arm 
of the beam leaving $BS_{1}$ is sent to $BS_{2}$ as an input, leading to 
the second part of the experiment where the simultaneous  
measurements of $\cos\phi$ and $\sin\phi$ are realized independently from 
the first part. $PBS_{2}$  
is aligned at a $45^{\circ}$ angle with respect to the reference  
frame selected by $PBS_{1}$. The intensities measured at the  
detectors $D_3$ and $D_4$ yield the measured values of $\cos{\phi}$  
and its moments. For the $sin\phi$ measurement, the phase of the  
remaining arm of the field is shifted by $\pi/2$ via a quarter 
wave-plate $\lambda/4$. The field is then     
sent to $PBS_{3}$ which is aligned in parallel to $PBS_{2}$. A 
simple calculation shows that the classical intensities measured  
at all detectors $D_{i}~, (i=1,..,6)$ are given by
\begin{equation}
\begin{array}{rlll}
I_{1}&=&{1 \over 2} \langle E_{1}^{2} \rangle ~, \\
\\
I_{2}&=&{1 \over 2} \langle E_{2}^{2} \rangle ~, \\
\\
I_{3}&=&{1 \over 4}[\langle E_{1}^{2} \rangle
+\langle E_{2}^{2} \rangle
+2\sqrt{\langle E_{1}^{2} \rangle\langle E_{2}^{2} \rangle}\cos\phi]~, \\
\\
I_{4}&=&{1 \over 4}[\langle E_{1}^{2} \rangle
+\langle E_{2}^{2} \rangle
-2\sqrt{\langle E_{1}^{2} \rangle\langle E_{2}^{2} \rangle}\cos\phi]~, \\
\\
I_{5}&=&{1 \over 4}[\langle E_{1}^{2} \rangle
+\langle E_{2}^{2} \rangle
+2\sqrt{\langle E_{1}^{2} \rangle\langle E_{2}^{2} \rangle}\sin\phi]~, \\
\\
I_{6}&=&{1 \over 4}[\langle E_{1}^{2} \rangle
+\langle E_{2}^{2} \rangle
-2\sqrt{\langle E_{1}^{2} \rangle\langle E_{2}^{2} \rangle}\sin\phi]~. \\
\end{array}
\label{I.2}
\end{equation}
Eqs\,(\ref{I.1}) and (\ref{I.2}) imply that the classical Stokes parameters 
can be extracted operationally by measuring all field intensities 
$I_{i}~~(i=1,..,6)$. 
In terms of these intensities, the Stokes parameters are simply 
given by
\begin{equation}
\begin{array}{rlll}
s_{0}&=(I_{1}+I_{2})~,~\qquad \qquad   
s_{1}&=(I_{1}-I_{2})~, \\
\\
s_{2}&=(I_{3}-I_{4})~,~\qquad \qquad  
s_{3}&=(I_{5}-I_{6})~. 
\end{array}
\label{I.3}
\end{equation}
In [11], we parameterized the polarized field   
in terms of the functions  
\begin{equation}
\begin{array}{rlrlrl}
\cos\theta&=s_{1}/s_{0}~,\quad & \qquad   
\sin\theta&=\sqrt{s_{0}^{2}-s_{1}^{2}}/s_{0}~,
\\
\\
\cos\phi&=s_{2}/\sqrt{s_{2}^{2}+s_{3}^{2}}
~,\quad &\qquad   
\sin\phi&=s_{3}/\sqrt{s_{2}^{2}+s_{3}^{2}}~.  
\end{array}
\label{I.4}
\end{equation}
This particular choice of the parameters above    
proves to be very convenient in the quantum operational 
measurements on fully polarized light. 
They also lead us naturally to the Poincar\`e's geometric 
interpretation of polarization.\cite{12,13,14,15}
Here $\theta$ and $\phi$ are physical parameters as 
shown in Fig.\,(2). Their values are directly connected with the 
ellipsometry of the polarized field. From now on we will adopt this 
parameterization and study the polarization fluctuations in terms of the 
fluctuations of these observables.    

\subsection{The Quantum Measurement Scheme}
The classical description above is adequate when the field intensity is  
sufficiently high. The vacuum fields, which are not present in the   
classical approach, are necessary for the correct quantum description  
of the apparatus as well as the field observables.      

In Fig.\,(1), the field operators $\hat{d}_{1},\,\hat{d}_{2}$ at the 
output of $PBS_{1}$  
are related to the input field components $\hat{a}_{1}, \hat{a}_{2}$ as
\cite{11}
\begin{equation}
\hat{d}_{1}={1 \over \sqrt{2}}\,(r\,\hat{a}_{1}+t\,\hat{v}^{(1)}_{1}),\qquad 
\hat{d}_{2}={1 \over \sqrt{2}}\,(r\,\hat{a}_{2}+t\,\hat{v}^{(1)}_{2})
\label{I.5}
\end{equation}
where $r=i/\sqrt{2}$ and $t=1/\sqrt{2}$ are the field reflection 
and transmission coefficients, and $\hat{v}_{j}^{(1)}~~(j=1,2)$ are the 
polarized vacuum fields entering through the vacuum port of $BS_{1}$.  
If the measurement scheme in Fig.\,(1) is extended to include  
the photodetectors $D_{i}~~~(i=3,4,5,6)$, then $\phi$   
 measurements can be made compatible with a proper 
quantum treatment of all fields.  
The output fields of $PBS_{2}$ and $PBS_{3}$ at  
$D_3,\,D_4,\,D_5,\,D_6$ are given by
\begin{equation}
\begin{array}{rlrl}
\hat d_{3}&={1 \over \sqrt{2}}[(t r \hat a_{1}+r^2   
\hat v^{(1)}_{1}+
t  \hat v^{(2)}_{1})+(t r \hat a_{2}+r^2   
\hat v^{(1)}_{2}+t  
\hat v^{(2)}_{2})]~&,\\  
\\
\hat d_{4}&={1 \over \sqrt{2}}[-(t r \hat a_{1}+r^2   \hat 
v^{(1)}_{1}+
t  \hat v^{(2)}_{1})+(t r \hat a_{2}+r^2   \hat v^{(1)}_{2}+
t  
\hat v^{(2)}_{2})]~&, \\
\\
\hat d_{5}&={1 \over \sqrt{2}}[i(t^{2} \hat a_{1}+t r  
\hat v^{(1)}_{1}+
r  \hat v^{(2)}_{1})+(t^{2} \hat a_{2}+t r  
\hat v^{(1)}_{2}+r  
\hat v^{(2)}_{2})]~&, \\ 
\\
\hat d_{6}&={1 \over \sqrt{2}}[-i(t^{2} \hat a_{1}+t r  
\hat v^{(1)}_{1}+
r  \hat v^{(2)}_{1})+(t^{2} \hat a_{2}+t r  
\hat v^{(1)}_{2}+r  
\hat v^{(2)}_{2})]~&.  
\end{array}
\label{I.6}
\end{equation}
In connection with their classical counterparts in Eqs\,(\ref{I.3}),  
we are now at a point to suggest the quantum Stokes 
parameters for the field operators $\hat{d}_{i}$ within this operational  
approach in terms of the observable photon number operators 
$\hat{n}_{i}=\hat{d}_{i}^{\dagger}\,\hat{d}_{i}$ as
\begin{equation}
\begin{array}{rlll}
\hat{\Sigma}_{0}=&\hat{n}_{1}+\hat{n}_{2}~, \qquad \qquad 
\hat{\Sigma}_{1}=&\hat{n}_{1}-\hat{n}_{2}~, \\
\\
\hat{\Sigma}_{2}=&\hat{n}_{3}-\hat{n}_{4}~, \qquad \qquad  
\hat{\Sigma}_{3}=&\hat{n}_{5}-\hat{n}_{6}~. \\
\end{array}
\label{I.7}
\end{equation}
In Eqs\,(6), all field operators commute as a manifestation of the 
vacuum fields. Hence in Eqs\,(\ref{I.7}) we have 
$[\hat{\Sigma}_{i},\hat{\Sigma}_{j}]=0~~(i \ne j)$ and all photon 
number operators can be simultaneously measured at the detectors 
$D_{i}~~(i=1,\dots,6)$. 
As a result, Eqs\,(\ref{I.7}) are compatible with their classical 
counterparts 
in Eqs\,(\ref{I.3}). This property of the 
$\hat{\Sigma}_{i}~,~~(i=0 \dots 3)$ 
operators allows us to further suggest an extension Eqs\,(\ref{I.4}) 
to their operator counterparts as
\begin{equation} 
\begin{array}{rlrl}
\hat{C}_{\theta}=&\hat{\Sigma}_{0}^{-1}\,\hat{\Sigma}_{1}~, \qquad \qquad
\qquad \qquad \,
\hat{S}_{\theta}=&(1-\hat{C}_{\theta}^{2})^{1/2} \qquad {\rm and} \\
\\
\hat{C}_{\phi}=&\hat{\Sigma}_{2}\,(\hat{\Sigma}_{2}^{2}+
\hat{\Sigma}_{3}^{2})^{-1/2}~, \qquad \qquad
\hat{S}_{\phi}=&\hat{\Sigma}_{3}\,(\hat{\Sigma}_{2}^{2}+
\hat{\Sigma}_{3}^{2})^{-1/2}~.
\end{array}
\label{I.8}
\end{equation}
$\hat{C}_{\theta}, \hat{S}_{\theta}$ as well as   
$\hat{C}_{\phi}, \hat{S}_{\phi}$ are well-defined and compatible quantum 
observables. They commute with each other and satisfy the operator 
relations $\hat{C}_{\theta}^{2}+\hat{S}_{\theta}^{2}=1~$, 
$~\hat{C}_{\phi}^{2}+\hat{S}_{\phi}^{2}=1~$ and, as the result, they can be 
measured simultaneously.

One of the benefits of adopting Eqs\,(\ref{I.7}) and (\ref{I.8}) 
is that all measurements are 
now based on pure photon counting depending on the measured photocounts  
at the detectors $D_{i}~,~~(i=1,\dots 6)$, and hence they do not    
involve any temporal interference effects. This is an advantage of the 
operational measurement, which  
will be transparent later in our discussion of the weak-field limit.   

Eqs\,(\ref{I.8}), hereinafter referred to as the  
{\it operational quantum Stokes parameters} (OQSP), are the most  
convenient choice for $\hat{\Sigma}_{i}~~(i=0,..,3)$   
befitting the purpose of the photocount measurement  
scheme of Fig.\,(\ref{I.1}). 
All operators in (\ref{I.8}) are now compatible with the classical 
variables of 
Eqs\,(\ref{I.4}) as long as the measurements of the $\hat{\Sigma}_{2}$ 
and $\hat{\Sigma}_{3}$ operators do not yield zero simultaneously.   

\section{The Measurement of Polarization Fluctuations in Weak Fields}
The operational approach as applied to the polarization measurement of 
a fully polarized and weak initial field 
is based on individual detections of single photons where the quantum   
nature of the field as well as of the detection mechanisms is dominant.  
The influence  
of the direct quantum homodyne detection on the statistics of a quantum 
measurement has been examined by Mandel\cite{16}, Kelley and Kleiner
\cite{17} as well as by Glauber\cite{18} and expressed 
for in the form of a combined quantum probability distribution 
\begin{equation}
{\cal P}(\{n_{j}\})=\prod_{j=1}^{N}~
:(\hat d_{j}^{\dagger} \hat d_{j})^{n_{j}}~exp(-\hat d_{j}^{\dagger}
\hat d_{j})/n_{j}!:~~~.
\label{meas.1}
\end{equation}
where $:~~:$ accounts for the normal ordering of the operators 
$\hat{d}_{i}^{\dagger}, \hat{d}_{i}$ inside and $\hat{d}_{i}^{\dagger}\,
\hat{d}_{i}$ corresponds to the photon number operator.  
Throughout the calculations the measurement time interval will be 
assumed to be much smaller than the coherence time (which is naturally 
satisfied for a monochromatic  field) and much larger than the inverse of 
the oscillation frequency of the field. Under these conditions 
it is possible to consider the simplest case when the photocount measurement  
at the detectors are time translationally invariant  
and linearly dependent on the measurement time interval $T$. 

Including the quantum effects of the homodyne detection in 
Eq.\,(\ref{meas.1}), an individual measurement of an arbitrary field 
operator $f(\{\hat{n}_{j}\})$ yields the measured value  
\begin{equation}
\langle f(\{\hat{n}_{j}\}) \rangle={\cal N}\,\sum_{\{n_{j}\}}\,
f(\{n_{j}\})\,Tr\Bigl\{\,\hat{\rho}\,{\cal P}(\{n_{j}\})\Bigr\}~~\qquad 
j=1,2,\dots M
\label{meas.2}
\end{equation}
where the trace is considered over the complete set of states in the 
density matrix of the initial field $\hat{\rho}=\vert \psi\rangle_{in} ~ 
_{in}\langle\psi \vert$. 

With Eqs\,(\ref{meas.1}) and (\ref{meas.2}) representing a general 
scheme of measurement in the operational approach, we now consider for  
$f(\{\hat{n}_{j}\})$ the operators of $\{\hat{n}_{j}\}~~,(j=1,2~
{\rm or}~3,4,5,6)$  
\begin{equation}
\hat{\cal E}_{\theta}(x)=(\hat{C}_{\theta}+i\,\hat{S}_{\theta})^{x}~,\qquad
{\rm and}\qquad
\hat{\cal E}_{\phi}(x)=(\hat{C}_{\phi}+i\,\hat{S}_{\phi})^{x} 
\qquad {\rm where} \quad x \in \mbox{\ee R}~. 
\label{meas.3}
\end{equation}
In the construction of $\hat{C}_{\theta}, \hat{S}_{\theta}$ and 
$\hat{C}_{\phi}, \hat{S}_{\phi}$ pairs in Eqs\,(\ref{I.8}), 
the compatibility 
conditions $[\hat{\Sigma}_{i}, \hat{\Sigma}_{j}]=0$ of the OQSP ensure 
that $\Vert \hat{\cal E}_{\theta}(x) \Vert=1$ and 
$\Vert \hat{\cal E}_{\phi}(x) \Vert=1$, hence $\hat{\cal E}_{\theta}(x)$ 
and $\hat{\cal E}_{\phi}(x)$ are unitary operators for all 
$x \in \mbox{\ee R}$.  
According to the procedure outlined in the context of Eqs\,(\ref{meas.1}) 
and (\ref{meas.2}), the measurements of these operators yield 
\begin{equation}
\langle \hat{\cal E}_{\theta}(x) \rangle={\cal N}_{\theta}\,
\sum_{\{n_{j}\}}\,\Bigl[\,\frac{n_{1}-n_{2}+2 i \sqrt{n_{1} n_{2}}}
{n_{1}+n_{2}}\Bigr]^{x}\,\langle {\cal P}(\{n_{j}\}) \rangle
\label{meas.4}
\end{equation}
and
\begin{equation}
\langle \hat{\cal E}_{\phi}(x) \rangle={\cal N}_{\phi}\,
\sum_{\{n_{j}\}}\,\Bigl[\frac{(n_3-n_4)+i(n_5-n_6)}
{\sqrt{(n_3-n_4)^2+(n_5-n_6)^2}}\Bigr]^{x}\,
\langle {\cal P}(\{n_{j}\}) \rangle
\label{meas.5}
\end{equation}
where $\langle {\cal P}(\{n_{j}\}) \rangle=Tr\Bigl\{ \, 
\vert \psi\rangle_{in}~_{in}\langle \psi \vert \, 
{\cal P}(\{n_{j}\})\Bigr\}$. In Eq.\,(\ref{meas.4}) $\{n_{j}\}=(n_1,n_2)$ 
and in Eq.\,(\ref{meas.5}) $\{n_{j}\}=(n_3,n_4,n_5,n_6)$. 
Clearly, Eq.(\ref{meas.4}) is well defined if  
$n_1, n_2$ are not simultaneously zero and, similarly, Eq.(\ref{meas.5})   
is well defined if $(n_3-n_4)$ and $(n_5-n_6)$ are not simultaneously zero 
in the respective summations above. 
The idea of the elimination of the configurations $n_1=n_2=0$ and $(n_3=n_4),
(n_5=n_6)$ from the statistical weight has been introduced as a crucial
element of the operational approach\cite{2,3,6,9} in the implementation of
the statistical averages. The effective weight of such configurations becomes
non-negligible particularly in the case when the initial field strength 
is sufficiently weak where the
probability of receiving zero number of photons within the detector's
measurement time interval $T$ is finite.
For instance, the weight of observing zero photons simultaneously
at the detectors $D_{1}, D_{2}$ is given by $\langle {\cal P}(0,0) \rangle$.
The result of such a null measurement is inconclusive in the calculation of
the averages in Eq.\,(\ref{meas.4}). Similarly, 
$n_{3}=n_{4}$ and $n_{5}=n_{6}$ yield
additional inconclusive results in the measurement on
$\langle \hat{\cal E}_{\phi}(x) \rangle$ in Eq.\,(\ref{meas.5}). The 
measured averages are then normalized by excluding the total
statistical weight of these inconclusive configurations from the integrated 
probability.
For strong fields, the weight of such ambiguous outcomes
is smaller and in the classical field limit there is no contribution from
such terms, viz., ${\cal N}_{\theta}={\cal N}_{\phi}=1$. In the 
measurement of the temporal phase the individual fluctuations
of these weak components as well as the fluctuations in the
relative number of photons can be strong due to the absence of a classical
reference source (i.e. a strong local oscillator).
Hence the normalization technique introduced by NFM proves to be
essential for any operational measurement based on phase and 
thus also for our approach here.

More explicitly, this normalization procedure amounts to\cite{2}  
\begin{equation}
{\cal N}_{\theta}^{-1}=1-\langle {\cal P}(0,0) \rangle 
\label{meas.5.1}
\end{equation}
and 
\begin{equation}
{\cal N}_{\phi}^{-1}=1-\sum_{n,m}\,
\langle {\cal P}(n_3=n_4=n, n_5=n_6=m) \rangle
\label{meas.5.2}
\end{equation}
in Eqs\,(\ref{meas.4}) and (\ref{meas.5}). 
The observed unitarity conditions of $\hat{\cal E}_{\theta}(x)$ 
and $\hat{\cal E}_{\phi}(x)$ 
suggest that one can associate a classical random variable 
$e^{ix \theta}$ and $e^{ix \phi}$ respecting the   
probability distributions $P(\theta)$ and $P(\phi)$  
such that\cite{6} 
\begin{equation}
\langle \hat{\cal E}_{\theta}(x)\rangle=\int_{0}^{\pi}\, 
\,d\theta\,
e^{ix\,\theta}\,
P(\theta)~,\qquad {\rm and} \qquad 
\langle \hat{\cal E}_{\phi}(x)\rangle=\int_{-\pi}^{\pi}\,
\,d\phi\,
e^{ix\,\phi}\,
P(\phi)~. 
\label{meas.10}
\end{equation} 
The probability distributions can then be obtained by the inverse 
Fourier transformations of (\ref{meas.10}) by 
\begin{equation}
P(\theta)=\int_{-\infty}^{\infty} \, 
\frac{dx}{2\pi}\,e^{-ix\,\theta}\,\Bigl\{
\langle \hat{\cal E}_{\theta}(x)\rangle+ 
\langle \hat{\cal E}_{\theta}(-x)\rangle \Bigr\} 
~, \qquad {\rm and} 
\qquad
P(\phi)=\int_{-\infty}^{\infty} \,
\frac{dx}{2\pi}\,e^{-ix\,\phi}\,
\langle \hat{\cal E}_{\phi}(x)\rangle 
\label{meas.11}
\end{equation}
with $\qquad \int_{0}^{\pi}\,d\,\theta
\,P(\theta)=\int_{-\pi}^{\pi}\,d\phi
\,P(\phi)=1$. 

Defining two auxiliary functions of $\{n\}_{j}$ by
\begin{equation}
\theta_{\{n\}}=tan^{-1}\Bigl(\frac{2\sqrt{n_1\,n_2}}{n_1-n_2}\Bigr)~,
\qquad {\rm and} \qquad
\phi_{\{n\}}=tan^{-1}\Bigl(\frac{n_5-n_6}{n_3-n_4}\Bigr)
\label{meas.6}
\end{equation}
where $\{n\}=(n_1,n_2)$ and $\{n\}=(n_3,n_4,n_5,n_6)$ for
$\theta$ and $\phi$ respectively, and 
using Eqs\,(\ref{meas.4}) and (\ref{meas.5}), the moments for a  
generalized initial state $\vert \psi\rangle_{in}$ read 
\begin{equation}
\begin{array}{rl}
\langle \hat{\cal E}_{\theta}(x) \rangle=&
{\cal N}_{\theta}\,\sum_{n_{1},n_{2}}^{\prime}\,
e^{ix\,\theta_{\{n\}}}\,\langle {\cal P}(\{n_{j}\})\rangle \, \qquad 
{\rm where}\\
\\
\langle {\cal P}(\{n_{j}\})\rangle=
&{\cal N}_{\theta}\,\sum_{n_{1},n_{2}}^{\prime}\,
\Bigl[\frac{(n_1-n_2)+i2\sqrt{n_1\,n_2}}{n_1+n_2}\Bigr]^{x}  \,
\frac{1}{2^{n_1+n_2}\,n_1! n_2!}
\\
&\times
\,_{in}\langle \psi\vert\,:\,(\hat{a}_{1}^{\dagger}\hat{a}_{1})^{n_1}\,
(\hat{a}_{2}^{\dagger}\hat{a}_{2})^{n_2}\,
exp[-1/2(\hat{a}_{1}^{\dagger}\hat{a}_{1}+
\hat{a}_{2}^{\dagger}\hat{a}_{2})]\,:\,\vert \psi\rangle_{in}\,
\end{array}
\label{meas.12}
\end{equation}
and
\begin{equation}
\begin{array}{rl}
\langle \hat{\cal E}_{\phi}(x; \delta_{0}) \rangle=&
{\cal N}_{\phi}\,\sum_{n_3,n_4,n_5,n_6}^{\prime}~
e^{ix\,\phi_{\{n\}}}\,
\langle {\cal P}(\{n_{j}\},\delta_{0})\rangle \qquad {\rm where} 
\\
\\
{\cal P}(\{n_{j}\},\delta_{0}) \rangle=&
{\cal N}_{\phi}\,\sum_{n_3,n_4,n_5,n_6}^{\prime}~ 
\Bigl[\frac{(n_3-n_4)+i(n_5-n_6)}
{\sqrt{(n_3-n_4)^2+(n_5-n_6)^2}}\Bigr]^{x} 
\,\frac{1}{8^{n_3+n_4+n_5+n_6}\,n_3! n_4! n_5! n_6!}
\\
&\times\,
_{in}\langle \psi \vert\,:\,
(\hat{a}^{\dagger}_{1}+\hat{a}^{\dagger}_{2})^{n_3}
\,(\hat{a}_{1}+\hat{a}_{2})^{n_3}\,
(-\hat{a}^{\dagger}_{1}+\hat{a}^{\dagger}_{2})^{n_4}\,
(-\hat{a}_{1}+\hat{a}_{2})^{n_4}\, \\
&\times\,
(-i\hat{a}^{\dagger}_{1}+\hat{a}^{\dagger}_{2})^{n_5}\,
(i\hat{a}_{1}+\hat{a}_{2})^{n_5}\,
(i\hat{a}^{\dagger}_{1}+\hat{a}^{\dagger}_{2})^{n_6}\,
(-i\hat{a}_{1}+\hat{a}_{2})^{n_6}\, \\
&\times \,exp[-1/2(\hat{a}_{1}^{\dagger}\hat{a}_{1}+
\hat{a}_{2}^{\dagger}\hat{a}_{2})]\,:\,\vert \psi\rangle_{in}\,~~~. 
\end{array}
\label{meas.13}
\end{equation}
with $\delta_{0}=\delta_{2}-\delta_{1}$ implicitly described in 
Eq.\,(\ref{meas.13}) as the relative temporal 
phase between the components of the initial field. The primes on the 
summations in Eqs\,(\ref{meas.12}) and (\ref{meas.13}) now indicate that 
the summations are performed by excluding those configurations for which the 
outcome is inconclusive. 

All moments are now determined once the initial components   
$\langle \hat{n}_{1}\rangle, \langle \hat{n}_{2}\rangle$ and the relative 
temporal phase $\delta_{0}$ of the inclusive fields 
$\hat{a}_{1}, \hat{a}_{2}$ are known. In our calculations the initial field 
parameters are chosen as the ratio of the photon numbers  
$\eta=\langle \hat{n}_{1}\rangle/\langle \hat{n}_{2}\rangle$, the total 
number of photons $\langle \hat{\Sigma}_{0}\rangle=
\langle \hat{n}_{1}\rangle+\langle \hat{n}_{2}\rangle$, and the relative 
temporal phase $\delta_{0}$.  

The credibility of the results obtained from the quantum operational 
approach crucially depends on the understanding of the influence of the 
quantum detectors on the final statistics. As pointed out in Refs.\,[2,3,6,9] 
that another essential element of the operational approach is  
the construction of an ensemble from a long series of such single 
operational measurements. 
The final physical results are then obtained by averaging the outcomes of 
single measurements over the created ensemble. Based on this 
prescription, we must now construct a physical ensemble  
of measured configurations in the calculations of the moments  
as well as the probability distributions in Eqs\,(\ref{meas.12}) and 
(\ref{meas.13}). The response of the quantum 
detectors to the incoming photons in the creation of the photocurrent 
is a random process which obeys the Poisson statistics in 
Eq.\,(\ref{meas.1}).\cite{19} As the photoelectrons are emitted at 
random times 
respecting this statistics, the information regarding the initial temporal 
phase $\delta_{0}$ of the incoming photons is modified and 
each repeated measurement is equivalent to superposing a random phase shift 
$\Delta$ on $\delta_{0}$.  
Hence the process of repeated measurements creates an ensemble of 
temporal phase configurations $\delta_{0}+\Delta$ 
with $\Delta$ being uniformly distributed over the available range. 
Since we consider in our calculations that the measurement time interval 
$T$ is considerably larger that the coherence time, the available 
range for $\Delta$ is the entire $2\pi$ range. Hence  
the average over the created ensemble corresponds to an averaging over  
a uniform distribution of $\Delta$. 
It is clear from Eq.\,(\ref{meas.12}) that the moments 
$\langle \hat{\cal E}_{\theta}(x)\rangle$ are independent from $\delta_{0}$; 
 hence they will also be independent of $\Delta$. This implies 
that a uniform average over $\Delta$ does not influence the measured moments 
$\langle \hat{\cal E}_{\theta}(x)\rangle$ and the probability distribution 
for $P(\theta)$ is given by 
\begin{equation}
P(\theta)=\int_{-\infty}^{\infty}\,\frac{dx}{2\pi}\,
e^{-ix\,\theta}\,\Bigl\{ 
\langle \hat{\cal E}_{\theta}(x)\rangle+ 
\langle \hat{\cal E}_{\theta}(-x)\rangle \Bigr\} \,   
~,\qquad 0 \le \theta \le \pi
\label{meas.14}
\end{equation}
On the other hand, the moments $\langle \hat{\cal E}_{\phi}(x)\rangle$ 
depend on the temporal phase $\delta_{0}$ and 
before the $\Delta$ average, $\delta_{0}$ dependence 
must be replaced by $\delta_{0}+\Delta$. This produces, at 
each measurement, the conditioned $\phi$ moments 
$\langle \hat{\cal E}_{\phi}(x;\delta_{0}+\Delta)\rangle$ 
and, following Ref.\,[6], their  
conditional probability distribution 
$P(\phi,\delta_{0};\Delta)$ is given by 
\begin{equation}
P(\phi,\delta_{0};\Delta)=\int_{-\infty}^{\infty}\,\frac{dx}{2\pi}\, 
\langle \hat{\cal E}_{\phi}(x;\delta_{0}+\Delta)\rangle\,
e^{-i\,x\,(\phi-\Delta)}~,\qquad -\pi \le \phi \le \pi
\label{meas.15}
\end{equation}
Therefore, the ensemble averaged probability distribution is
\begin{equation}
P(\phi,\delta_{0})=
\int_{-\pi}^{\pi}\,\frac{d\Delta}{2\pi}\,P(\phi,\delta_{0};\Delta)~.
\label{meas.16}
\end{equation}
After a short calculation using Eqs\,(\ref{meas.12}) and (\ref{meas.13}) 
in Eqs.\,(\ref{meas.14}-\ref{meas.16}),  
the probability 
distributions $P(\theta)$ and $P(\phi,\delta_0)$ can be 
expressed by  
\begin{equation}
P(\theta)={\cal N}_{\theta}\,\sum_{\{n_{j}\}}\,
\delta(\theta-\theta_{n})\,
\langle {\cal P}(\{n_{j}\})\rangle
\label{meas.14.b}
\end{equation}
and
\begin{equation}
P(\phi,\delta_{0})={\cal N}_{\phi}\sum_{\{n_{j}\}}\,
\langle {\cal P}(\{n_{j}\}, \delta_{0}-\phi_{n}+\phi)
\rangle~.
\label{meas.16.b}
\end{equation}
where the last term in Eq.\,(\ref{meas.16.b}) is obtained by using 
Eqs\,(\ref{meas.13}) in Eq.\,(\ref{meas.15}). 

On the other hand, the detectors' influence on the measured statistics 
can only be understood if the measured moments and probability distributions 
are compared with those without the detectors' influence. 
For this purpose and, following Refs.\,[2,3], we define the theoretically 
inferred values of the $\theta$ and $\phi$ moments as 
$\langle \hat{\cal E}^{I}_{\theta}(x)\rangle$ and    
$\langle \hat{\cal E}^{I}_{\phi}(x)\rangle$ where 
\begin{equation}
\langle \hat{\cal E}^{I}_{\theta}(x)\rangle=_{in}\langle \psi \vert :\Bigl[ 
\frac{\hat{n}_{1}-\hat{n}_{2}+i\,2\,
\sqrt{\hat{n}_{1}\,\hat{n}_{2}}}
{\hat{n}_{1}+\hat{n}_{2}}\Bigr]^{x}:
\vert \psi \rangle_{in}
\label{meas.17}
\end{equation}
and
\begin{equation}
\langle \hat{\cal E}^{I}_{\phi}(x)\rangle=_{in}\langle \psi \vert :\Bigl[
\frac{(\hat{n}_{3}-\hat{n}_{4})+i\,(\hat{n}_{5}-\hat{n}_{6})}
{\sqrt{(\hat{n}_{3}-\hat{n}_{4})^{2}+
(\hat{n}_{5}-\hat{n}_{6})^{2}}}\,\Bigr]^{x}:\vert \psi \rangle_{in}
\label{meas.18}
\end{equation}
where $: ~~ :$ stands for the normal ordering of the field and vacuum 
operators inside. 

We calculate the probability distributions 
$P(\theta)$, where $0 \le \theta \le \pi$, and  
$P(\phi,\delta_0)$, where $-\pi \le \phi \le \pi$    
numerically using Eqs\,(\ref{meas.14.b}) and (\ref{meas.16.b}). 
Since $\cos{\theta}$ 
is single valued in the $\theta$ range considered, we will only  
need to examine the fluctuations in the $\hat{C}_{\theta}$ operator. On the 
other hand, in the $\phi$ range considered both $\hat{C}_{\phi}$ 
and $\hat{S}_{\phi}$ operators will be necessary. In our calculations, 
the summations over infinite range of $\{n\}_{j}$'s are truncated at 
$\{n\}_{j}^{max}=20$ for all $j$ which naturally restrict the accuracy of 
the results to sufficiently weak initial fields. 
The measured moments and the probability distributions are then compared 
with the theoretically inferred ones by using Eqs\,(\ref{meas.17}) and 
(\ref{meas.18}). 

\subsection{Calculations for a fully polarized quantum coherent field:}
Let us now assume that the initial field is in a fully polarized 
quantum coherent state  
$\vert \psi\rangle_{in}=\vert \alpha_{1}, \alpha_{2} \rangle$, with the 
parameters given by 
$\alpha_{j}=\vert \alpha_{j} \vert\,e^{i\delta_{j}}$ where 
$\vert \alpha_{j}\vert^{2}$ and $\delta_{j}~~~(j=1,2)$ are the average  
number of photons and the coherent temporal 
phase of the $j$'th component respectively. The relative temporal 
phase is given, as before, by $\delta_{0}=\delta_{2}-\delta_{1}$. 
From Eqs\,(\ref{meas.12}) and 
(\ref{meas.13}), the measured moments in this state are given by  
\begin{equation}
\begin{array}{rl}
\langle \hat{\cal E}_{\theta}(x) \rangle=&
{\cal N}_{\theta}\,\sum_{n_{1},n_{2}}^{\prime}\, 
\Bigl[\frac{(n_1-n_2)+i2\sqrt{n_1\,n_2}}{n_1+n_2}\Bigr]^{x} \\
\\
&\times \vert \alpha_{1} \vert^{2n_1}\,
\vert \alpha_{2} \vert^{2n_2}\,exp\{-\frac{1}{2}(\vert \alpha_{1}\vert^{2}+
\vert \alpha_{2}\vert^{2})\}/2^{n_1+n_2}\,n_1! n_2!~~~,
\end{array}
\label{coh.1}
\end{equation}
and 
\begin{equation}
\begin{array}{rl}
\langle \hat{\cal E}_{\phi}(x; \delta_{0}) \rangle=&
{\cal N}_{\phi}\,\sum_{\{n\}}^{\prime}~ 
\Bigl[\frac{(n_3-n_4)+i(n_5-n_6)}{\sqrt{(n_3-n_4)^2+(n_5-n_6)^2}}\Bigr]^{x} \\
\\
&\times 
\vert \alpha_1+\alpha_2\vert^{2n_3}\,
\vert -\alpha_1+\alpha_2\vert^{2n_4}\,
\vert -i \alpha_1+\alpha_2\vert^{2n_5}\,
\vert i \alpha_1+\alpha_2\vert^{2n_6} \\ 
\\
&\times 
exp\{-\frac{1}{2}(\vert \alpha_{1}\vert^{2}+
\vert \alpha_{2}\vert^{2})\}/8^{n_3+n_4+n_5+n_6}\,n_3! n_4! n_5! n_6!~~~,  
\end{array}
\label{coh.2}
\end{equation}
where $\{ n \}=(n_3,n_4,n_5,n_6)$. For the specific initial polarized  
coherent state considered, using Eqs(\ref{meas.5.1}) and 
(\ref{meas.5.2}), the normalizations are given by
\begin{equation}
{\cal N}_{\theta}^{-1}=1-exp\{-\frac{1}{2}(\vert \alpha_{1}\vert^{2}+
\vert \alpha_{2}\vert^{2})\}
\label{meas.17.1}
\end{equation}
and, defining $\beta=2\vert \alpha_{1}\vert \, \vert \alpha_{2}\vert/
(\vert \alpha_{1}\vert^{2}+\vert \alpha_{2}\vert^{2})$ where $\beta \le 1$, 
\begin{equation}
\begin{array}{rl}
{\cal N}_{\phi}^{-1}=1-\sum_{n,m}\,\Bigl(
\frac{\vert \alpha_{1}\vert^{2}+\vert \alpha_{2}\vert^{2}}{8}\Bigr)^{2(n+m)}\,
&\frac{exp\{-\frac{1}{2}(\vert \alpha_{1}\vert+\vert \alpha_{2}\vert^{2})\}}
{(n!)^{2}\,(m!)^{2}}. \\
& \times \Bigl[1-\beta^{2}\,\cos^{2}\delta_{0}\Bigr]^{n}\,
\Bigl[1-\beta^{2}\,\sin^{2}\delta_{0}\Bigr]^{m}
\end{array}
\label{meas.17.2}
\end{equation}
We will first examine the $P(\theta)$ distribution. Using  
Eq.\,(\ref{meas.14.b}) the calculation of $P(\theta)$ yields 
\begin{equation}
P(\theta)={\cal N}_{\theta}\,
\sum_{\{n_{j}\}, j=1,2}^{\prime}\,\delta(\theta-
\theta_{\{n\}})\,\frac{\vert \alpha_{1}\vert^{2n_1}\,
\vert \alpha_{2} \vert^{2n_2}}{2^{n_1+n_2}\,n_1! n_2!}
\,exp\{-\frac{1}{2}(\vert \alpha_{1}\vert^{2}+
\vert \alpha_{2}\vert^{2})\}
\label{coh.4}
\end{equation}
where $\theta_{\{n\}}$ is defined by the first expression in 
Eqs\,(\ref{meas.6}). 
For sufficiently weak fields, i.e., $\langle \Sigma_{0}\rangle \ll 1$, 
each detector measures null or a very few number of photons. This implies 
that in Eqs\,({\ref{coh.4})  
it is sufficient to restrict the summation over 
$\{n_{j}\}~,(j=1,2)$ to a few terms. For instance, let us consider 
$\{n_{j}\}=0,1$. Then including only the first-order 
terms in the average total photon number, Eq.\,(\ref{coh.4}) 
can be approximately expressed in the weak-field limit by 
$P_{w}(\theta)$ in the form  
\begin{equation} 
P_{w}(\theta)={\cal N}_{w}\,
\langle \Sigma_{0}\rangle\,\Biggl\{\,
\frac{1}{1+\eta^{-1}}\,\delta(\theta)+
\frac{1}{1+\eta}\,\delta(\theta-\pi)\,\Biggr\}
\label{coh.5}
\end{equation}
where $\langle \Sigma_{0}\rangle=(|\alpha_{1}|^{2}+|\alpha_{2}|^{2})/2$ is 
the total average photon number deduced from the 
measurements at the detectors $D_{1}, D_{2}$,  
$\eta=|\alpha_{1}|^{2}/|\alpha_{2}|^{2}$ and 
${\cal N}_{w}=\langle \Sigma_{0}\rangle^{-1}$ so that 
$\int_{0}^{\pi}\,d\theta\,P(\theta)=1$. 
From Eq.\,(\ref{coh.5}) we find that 
\begin{equation}
\langle \cos{\theta}\rangle_{w} \equiv 
\int_{0}^{\pi}\,d\theta\,\cos{\theta}\,P(\theta) =  
\frac{\eta-1}{\eta+1}~, \qquad 
\langle\cos^{2}{\theta}\rangle_{w}=1
\label{coh.6}
\end{equation}
Clearly, $\langle\cos{\theta}\rangle_{w}$ in Eq.\,(\ref{coh.6}) is 
consistent with the theoretically inferred values calculated from 
Eq.\,(\ref{meas.17}), [i.e. $\langle\cos{\theta}\rangle_{w}=  
\langle \hat{C}_{\theta}^{I}\rangle$]. In the initial 
polarized coherent state the theoretically inferred moments are given by 
\begin{equation}
\langle \hat{\cal E}^{I}_{\theta}(x)\rangle=
\Bigl[\frac{\eta-1+i\,2\,\eta^{1/2}}{\eta+1}\,\Bigr]^{x}=
e^{i\,x\,\tan^{-1}2\sqrt{\eta}/(\eta-1)}~,
\label{coh.7}
\end{equation}
which respect a nonfluctuating distribution. Equation (\ref{coh.7}) 
is also consistent with the classical calculations using Eq.\,(\ref{I.4}).
However for the second moments we obtain 
\begin{equation}
\langle (\hat{C}_{\theta}^{I})^{2}\rangle=
\Bigl(\frac{\eta-1}{\eta+1}\Bigr)^{2} \ne 
\langle\cos^{2}{\theta}\rangle_{w}~. 
\label{coh.6.1}
\end{equation}
The $\theta$ distribution in Eq.\,(\ref{coh.4}) is plotted in 
Fig.\,3 below for $\eta=1.0, 0.5$ and 
$\langle \hat{\Sigma}_{0}\rangle=0.1,1.0, 5.0, 10.0$. 
The first observation in Fig.\,3(a) is that at $\eta=1.0$ the probability
distribution is symmetrically centered around $\theta=\pi/2$. In the
weak-field limit $P(\theta)$ is peaked at
$\theta=0, \pi$. As the field strength is sufficiently increased, 
the central peak at $\theta=\pi/2$ gradually develops    
as all other peaks are suppressed. 
The average of the $\cos{\theta}$ within the full range 
$0 \le \theta \le \pi$ is zero 
as it would also be expected from the  
theoretically inferred moments in Eq.\,(\ref{coh.7}). 
For $\eta \ne 1$, the measured $P(\theta)$ is plotted in Fig.3(b). 
The delta functions in
Eq.\,(\ref{coh.4}) are numerically simulated by sharp Lorentzians, hence
they acquire a finite width in Figs.\,3(a) and 3(b).
On the other hand, using the equation in (\ref{meas.14}), the inferred 
probability distribution $P^{I}(\theta)$ can be found as 
$P^{I}(\theta)=
\delta\Bigl(\theta-\cos^{-1}(\eta-1)/(\eta+1)\Bigr)$. 

A similar calculation can also be done for the $P(\phi)$ 
distribution by making use of Eqs\,(\ref{coh.2}), (\ref{meas.15}) and 
(\ref{meas.16}). After some calculation using the normalization procedure 
leading to Eq.\,(\ref{meas.16}) we find that 
\begin{equation}
\begin{array}{rl}
P(\phi,\delta_{0})=
&{\cal N}_{\phi}\,\sum_{\{n_{j}\}~,(j=3,4,5,6)}^{\prime}\, 
\frac{e^{-\frac{1}{2}(\vert \alpha_{1}^{2}\vert+\vert \alpha_{2}\vert^{2})}}
{n_3!\,n_4!\,n_5!\,n_6!}\,\Bigl( \frac{\vert \alpha_{1}^{2}\vert+ 
\vert \alpha_{2}\vert^{2}}{8}\Bigr)^{n_3+n_4+n_5+n_6} \\
&\times [1+\beta\,\cos(\delta_{0}-\phi-
\phi_{\{n\}})]^{n_3}\,
[1-\beta\,\cos(\delta_{0}-\phi-
\phi_{\{n\}})]^{n_4}\,\\
&\times [1+\beta\,\sin(\delta_{0}-\phi-
\phi_{\{n\}})]^{n_5}\,
[1-\beta\,\sin(\delta_{0}-\phi-
\phi_{\{n\}})]^{n_6}
\end{array}
\label{coh.8}
\end{equation}
which is, not surprisingly, the same distribution obtained by NFM in 
Ref.\,[6] in a slightly different context. The weak-field limit of 
Eq.\,(\ref{coh.8}) has also been studied in 
Ref.\,[6] to where we refer the reader for additional details. The 
numerically calculated Eq.\,(\ref{coh.8}) is plotted in Fig.\,4  
for $\eta=1.0, 0.5$, $\Sigma_{0}=0.1,1.0,5.0,10.0$ and for 
$\delta_{0}=0$. The first observation 
we make here is that $P(\phi,\delta_{0})$ is almost independent from  
$\eta$ but strongly dependent on the strength of the   
initial field. As the field strength increases, the fluctuations decrease 
and the distribution becomes gradually narrower. On the other hand,  
using Eq.\,(\ref{meas.18}), the theoretically inferred moments are 
calculated as
\begin{equation}
\langle \hat{\cal E}^{I}_{\phi}(x)\rangle=e^{i\delta_{0}\,x}~. 
\label{coh.9}
\end{equation}
Hence, the theoretically inferred $\phi$ distribution is also 
non-fluctuating given by $P(\phi,\delta_{0})=
\delta(\phi-\delta_{0})$. 
The operational averages for $\phi$ 
as well as the probability distributions are strongly peaked 
in the strong-field limit and they have the tendency to approach to the 
theoretically inferred moments and the delta function-like probability 
distributions respectively. On the other hand, the operational approach 
predicts large deviations 
of the measurement from the theoretical values in the weak-field limit.  
In order to understand the influence of the photodetection particularly in 
the weak-field limit, we now examine the second order fluctuations in the 
measured moments of the $\theta$ and $\phi$ related operators. 

\subsubsection{The measured fluctuations in polarization}
Once the moments in Eqs\,(\ref{coh.1}) and (\ref{coh.2}) are defined, the 
measured moments of the cosine and sine operators of $\theta$ and $\phi$  
can be found. The same moments can also be found by the use of the 
probability distributions in Eqs\,(\ref{coh.4}) and (\ref{coh.8}). 
Here the weak-field limit is particularly interesting 
and it can also be examined analytically. We start our analysis of 
the fluctuations by reminding that, since $0 \le \theta \le \pi$, we will 
be confined to the measured fluctuations in the $\hat{C}_{\theta}$ 
operator. In the weak-field limit (keeping only the leading 
term in the total field strength) 
\begin{equation}
\langle \hat{C}_{\theta}\rangle \simeq \frac{\eta-1}{\eta+1}~,\qquad 
\langle \hat{C}_{\theta}^{2}\rangle \simeq 1~,  
\label{cohfl.1}
\end{equation} 
we find the dispersion $D(\theta)$ as 
\begin{equation}
D(\theta)=\sqrt{\langle (\hat{C}_{\theta}-\langle 
\hat{C}_{\theta} \rangle)^{2}\rangle} \simeq \frac{2 \eta^{1/2}}{\eta+1} 
\le 1 
\label{cohfl.2}
\end{equation}
The dependence of $D(\theta)$ for $\eta=1.0,0.5,0.1 $ 
on the total field strength  
is shown in Fig.\,5. 
We now shift our attention to the measured fluctuations in the 
$\phi$ distribution. Since $-\pi \le \phi \le \pi$ we  
need to consider here both the cosine and the sine moments.  
Considering first the weak-field limit, 
and keeping only linear terms in the total field strength, have,  
\begin{equation}
\langle \hat{C}_{\phi}\rangle \simeq 
\frac{\eta^{1/2}\, \cos{\delta_{0}}}{\eta+1}~, \qquad 
\langle \hat{S}_{\phi}\rangle \simeq 
\frac{\eta^{1/2}\, \sin{\delta_{0}}}{\eta+1} \qquad 
\langle \hat{C}^{2}_{\phi}\rangle=\langle \hat{S}^{2}_{\phi}\rangle=
\frac{1}{2}
\label{cohfl.3}
\end{equation}
In this case we define the dispersion $D(\phi)$ as
\begin{equation}
D(\phi)=\sqrt{\langle (\hat{C}_{\phi}-\langle 
\hat{C}_{\phi} \rangle)^{2}\rangle+
\langle (\hat{S}_{\phi}-\langle 
\hat{S}_{\phi} \rangle)^{2}\rangle}=\sqrt{1-\frac{\eta}{(\eta+1)^{2}}} \ge 
\frac{\sqrt{3}}{2}
\label{cohfl.4}
\end{equation}
The dependence of $D(\phi)$ on the total field strength is plotted 
for $\eta=1.0,0.5,0.1$ in Fig.\,6. 

On the other hand, for both the $\theta$ and the $\phi$ related moments the 
theoretically inferred fluctuations vanish in the coherent state 
(i.e. $D(\theta)^{I}=D(\phi)^{I}=0)$. 
The measured fluctuations differ significantly from 
the theoretically inferred ones as the strength of the initial field 
becomes weak. These deviations in Figs.5 and 6 
from the theoretically inferred values arise from the nature of the 
photodetection of weak fields and the normalization of the probability 
weight after 
discarding the inconclusive data. We find that the results for the fully 
polarized coherent field and the differences between the measured and 
inferred fluctuations in the weak-field limit closely relate to the 
results obtained by NFM.\cite{2,6} 

The $\eta$ dependence of the fluctuations in Figs.\,5 and 6 implies 
that $\eta$ can be used as a parameter in the measurement to search for 
an optimum orientation of the set-up in Fig.\,1 by rotating the 
reference axes 
$1,2$ around the initial field direction. Note that this corresponds to a 
solid rotation of the entire setup since the relative orientation of each  
polarizing beam splitter with respect to $PBS_{1}$ is fixed. By this 
operation the angle between the polarization axes of the setup and the 
mean principle axes of the polarization ellipse of the initial field can be 
changed. Let us suppose that $\alpha_{10}$ and $\alpha_{20}$ are initial 
coherent state parameters defined with respect to some fixed orientation 
$1_{0}, 2_{0}$ of $PBS_{1}$ and given by 
$\alpha_{j0}=\vert \alpha_{j0}\vert\,e^{i\delta_{j0}}$. 
 If the principle axes $1,2$ are rotated by an angle 
$\gamma$ with respect to $1_{0}, 2_{0}$, the initial coherent field 
parameters $\alpha_{1}, \alpha_{2}$ are effectively rotated by the same 
angle with respect to $\alpha_{10}$ and $\alpha_{20}$. In particular, 
the average number of photons $\langle \hat{n}_{1}\rangle$ and 
$\langle \hat{n}_{2}\rangle$ 
measured at $D_{1}$ and $D_{2}$ are given by
\begin{equation} 
\langle n_{1}\rangle=\frac{1}{2}\,\Bigl\vert \alpha_{10}\,\cos{\gamma}-
\alpha_{20}\,\sin{\gamma}\Bigr\vert^{2}~,\qquad {\rm and} \qquad 
\langle n_{2}\rangle=\frac{1}{2}\,\Bigl\vert \alpha_{20}\,\cos{\gamma}+
\alpha_{10}\,\sin{\gamma}\Bigr\vert^{2}
\label{cohfl.5}
\end{equation}
Since $\gamma$ is arbitrary, we can use it to tune 
$\eta=\langle \hat{n}_{1}\rangle/\langle \hat{n}_{2}\rangle$ in order 
to find whether an optimum orientation of the setup exists 
such that both  $\theta$ and $\phi$ related measurements 
(or whatever other observables are examined) can be {\it improved} 
simultaneously. 
The measured $\eta$ at the detectors $D_{1,2}$ is then a function of 
$\gamma$ and is represented in terms of the initially 
fixed $\eta_{0}=\vert \alpha_{10}\vert^{2}/\vert \alpha_{20}\vert^{2}$ 
and the relative phase $(\delta_{20}-\delta_{10})$ as
\begin{equation}
\eta(\gamma)=\frac{(\sqrt{\eta_0}-\tan{\gamma})^{2}+\sqrt{\eta_0}\,
\tan{\gamma}\,\sin^{2}{(\delta_{20}-\delta_{10})/2}}
{(\sqrt{\eta_0}+\tan{\gamma})^{2}-\sqrt{\eta_0}\,\tan{\gamma}\,
\sin^{2}{(\delta_{20}-\delta_{10})/2}}
\label{cohfl.6}
\end{equation}
The second order fluctuations represented by $D(\theta)$ and $D(\phi)$ 
are plotted in the weak-field limit as a function of $\gamma$ for 
$0 \le \gamma \le \pi/2$ and for $\langle \Sigma_{0}\rangle=1$ in 
Fig.7 and for $\langle\Sigma_{0}\rangle=9$ in Fig.8.  
The figures imply that such an optimum orientation to simultaneously 
minimize the fluctuations $D(\theta)$ and $D(\phi)$ 
for a fixed value of $\gamma$, 
$\delta_{20}-\delta_{10}$ and $\langle \Sigma_0\rangle$ does not exist. 
Hence, depending on the measured observable, one has to engineer such  
optimum configurations for each measurement independently. 

\subsection{Calculations for a fully polarized Fock state:}
Now let us assume that the initial field is given by the    
fully polarized photon 
number state $\vert \psi\rangle_{in}=\vert M \rangle_{\phi_0}$ 
as
\begin{equation}
\vert M \rangle_{\phi_{0}}={\cal N}_{M}\,\sum_{m=0}^{M}~
{M \choose m}^{1/2}~e^{i\phi_{0}\,(M-m)}~\vert m\,,\, M-m\rangle
\label{fock.1}
\end{equation}
where $\phi_{0}$ is a temporal phase between the polarization 
components, ${\cal N}_{M}=2^{-M/2}$,    
${M \choose m}$ is the binomial coefficient and 
$\vert m, M-m\rangle \equiv \vert m\rangle \otimes \vert M-m\rangle$ 
describes the relative number of photons in each component in 
reference to a particular choice of predefined axes of polarization.  
It is possible to see that the field operators for the Fock state 
\begin{equation}
\hat{a}_{\phi_0}=
\frac{1}{\sqrt{2}}\,(\hat{a}_{1}+e^{-i\phi_{0}}\,\hat{a}_{2})~,
\qquad {\rm and} \qquad 
\hat{a}^{\dagger}_{\phi_0}=\frac{1}{\sqrt{2}}\,(\hat{a}_{1}^{\dagger}+
e^{i\phi_{0}}\,\hat{a}_{2}^{\dagger})  
\label{fock.2}
\end{equation}
satisfy 
\begin{equation}
\hat{a}_{\phi_0}\,\vert M \rangle_{\phi_{0}}=\sqrt{M}\,
\vert M-1\rangle_{\phi_{0}}~, \qquad {\rm and} \qquad 
\hat{a}^{\dagger}_{\phi_0}\,\vert M \rangle_{\phi_{0}}=\sqrt{M+1}\,
\vert M+1\rangle_{\phi_{0}} 
\label{fock.3}
\end{equation}
The temporal phase factor $\phi_{0}$ 
determines the ellipticity of the polarization. 
If $\phi_{0}=0, \pi$, Eq.\,(\ref{fock.1}) provides the basis for linear 
polarization.    
For $\phi=\pm \pi/2$ left and right circularly polarized states 
are obtained. For arbitrary $\phi_{0}$ left and right elliptically polarized 
Fock states can be produced. Restricting $\phi_{0}$ within the range 
$0 \le \phi_{0} \le \pi$, the left and right elliptically polarized states 
are realized respectively by $\vert M\rangle_{\phi_{0}}$ and 
$\vert M\rangle_{(\phi_{0}-\pi)}$ with the respective field operators 
$\hat{a}_{\phi_0}, \hat{a}^{\dagger}_{\phi_0}$; 
$\hat{a}_{(\phi_0-\pi)}, \hat{a}^{\dagger}_{(\phi_0-\pi)}$. The second pair 
of field operators are then found by making the change 
$\phi_{0} \to \phi_{0}-\pi$ in Eqs\,(\ref{fock.2}). The field angular 
momentum operator is given by 
$\hat{\cal L}_{z}=(\hat{a}_{\phi_0}^{\dagger}\,\hat{a}_{\phi_0}-
\hat{a}_{\phi_0-\pi}^{\dagger}\,\hat{a}_{\phi_0-\pi})$ and $\hat{\cal L}_{z}$ 
is diagonal in $\vert M\rangle_{\phi_0}$ with eigenvalue $M$. 

Using Eqs\,(\ref{fock.2}) and (\ref{fock.3}),  Eq.\,(\ref{fock.1}) 
can be written as 
\begin{equation}
\vert M \rangle_{\phi_0}=\frac{1}{\sqrt{M!}}\,(\frac{\hat{a}_{1}^{\dagger}+
e^{i\phi_{0}}\hat{a}_{2}^{\dagger}}{\sqrt{2}})^{M}\,\vert 0 , 0\rangle
=\frac{1}{\sqrt{M!}}\,(\hat{a}_{\phi_0}^{\dagger})^{M}\,\vert 0\rangle
\label{fock.1.b}
\end{equation}
where $\vert 0\rangle$ is the vacuum state for $\hat{a}_{\phi_0}$ as 
well as for $\hat{a}_{1}$ and, $\hat{a}_{2}$; (i.e., 
$\vert 0\rangle \equiv \vert 0,0\rangle$). In what follows, the full 
range $-\pi \le \phi_{0} \le \pi$ will be considered. 
In fact, Eq\,(\ref{fock.1}) 
is an example in a class of fully polarized Fock states corresponding 
to $\eta=\langle \hat{n}_{1}\rangle/\langle \hat{n}_{2}\rangle=1$ where 
$\langle \hat{n}_{1}\rangle$ and $\langle \hat{n}_{2}\rangle$ describe 
average number photons in individual polarization modes. For 
Eq.\,(\ref{fock.1}) we have $\langle \hat{n}_{1}\rangle=
\langle \hat{n}_{2}\rangle=M/2$. If a rotation parameter $\gamma$ is 
introduced [for instance, like in Eq.\,(\ref{cohfl.5}) for the coherent 
state] in the field space by  
\begin{equation}
\hat{a}_{\phi_0,\gamma}=(\cos{\gamma}\,\hat{a}_{1}+
e^{-i\phi_{0}}\,\sin{\gamma}\,\hat{a}_{2})~,
\qquad {\rm and} \qquad
\hat{a}^{\dagger}_{\phi_0,\gamma}=(\cos{\gamma}\,\hat{a}_{1}^{\dagger}+
e^{i\phi_{0}}\,\sin{\gamma}\hat{a}_{2}^{\dagger})~, 
\label{fock2.b}
\end{equation}
in terms of the new field operators $\hat{a}_{\phi_0,\gamma},
\hat{a}^{\dagger}_{\phi_0,\gamma}$ the field operators of the initial 
Fock state in Eq.\,(\ref{fock.1}) would be obtained when   
$\gamma=\pi/4$ in Eq.\,(\ref{fock2.b}). This implies 
that the Fock state $\vert M\rangle_{\phi_0,\gamma}$ created by 
Eq.\,(\ref{fock2.b}) is realized effectively by a $(\gamma-\pi/4)$ degree  
rotation of the Fock state in Eq.\,(\ref{fock.1}) with 
$\vert M\rangle_{\phi_0,\gamma}$ being     
\begin{equation}
\vert M\rangle_{\phi_0,\gamma}=\frac{1}{\sqrt{M!}}\,
(\hat{a}_{\phi_0,\gamma}^{\dagger})^{M}\,\vert 0 \rangle=
\sum_{m=0}^{M}\,{M \choose m}^{1/2}\,(\cos{\gamma})^{m}\,
(e^{i\phi_0}\,\sin{\gamma})^{(M-m)}\,\vert m , M-m\rangle~.  
\label{fock1.c}
\end{equation}
Eq.\,(\ref{fock1.c}) for a fixed $\phi_0$ now describes a fully polarized 
generalized Fock state with an arbitrary 
ratio of photon numbers $\eta(\gamma)=\cot^{2}{\gamma}$ 
between the polarization components. 

In comparison with the coherent initial field, a 
considerably more tedious work is involved in the numerical calculations of 
both measured moments. In the general fully polarized Fock state 
given by Eq.\,(\ref{fock1.c}), Eqs\,(\ref{meas.4}) and (\ref{meas.5}) 
become  
\begin{equation}
\langle \hat{\cal E}_{\theta}(x)\rangle={\cal N}_{\theta}\,
\sum_{n_{1},n_{2}}^{\prime}\,
\Bigl[\frac{(n_1-n_2)+i2\sqrt{n_1\,n_2}}{n_1+n_2}\Bigr]^{x} \,
\frac{1}{2^{n_1+n_2}\,n_1!\,n_2!}\,
\Bigl\{\sum_{r,p}\,\frac{(-1)^{r+p}}{2^{r+p}\,r!\,p!}\,
\mu_{(n_1+r)}^{(n_2+p)}\,\Bigr\} 
\label{fock.4}
\end{equation}
and 
\begin{equation}
\begin{array}{rl}
&\langle \hat{\cal E}_{\phi}(x,\phi_0) \rangle =
{\cal N}_{\phi}\,\sum_{\{n\}}^{\prime}\,
\Bigl[\frac{(n_3-n_4)+i(n_5-n_6)}
{\sqrt{(n_3-n_4)^{2}+(n_5-n_6)^{2}}}\Bigr]^{x}\,
\frac{1}{8^{n_3+n_4+n_5+n_6}\,n_3!\,n_4!\,n_5!\,n_6!}\, \\
\\
&\times \sum_{r}\,\frac{(-1)^{r}}{2^{r}\,r!}\,
.\langle M \vert : (\hat{a}_{1}^{\dagger}\,\hat{a}_{1}+
\hat{a}_{2}^{\dagger}\,\hat{a}_{2})^{\tilde{n}+r} \\
\\
&\times \Bigl(1+\frac{\hat{a}_{1}^{\dagger} \hat{a}_{2}+
\hat{a}_{2}^{\dagger} \hat{a}_{1}}{\hat{a}_{1}^{\dagger} \hat{a}_{1}+
\hat{a}_{2}^{\dagger} \hat{a}_{2}}\Bigr)^{n_3}\,
\Bigl(1-\frac{\hat{a}_{1}^{\dagger} \hat{a}_{2}+
\hat{a}_{2}^{\dagger} \hat{a}_{1}}{\hat{a}_{1}^{\dagger} \hat{a}_{1}+
\hat{a}_{2}^{\dagger} \hat{a}_{2}}\Bigr)^{n_4}\,
\Bigl(1-i\frac{\hat{a}_{1}^{\dagger} \hat{a}_{2}-
\hat{a}_{2}^{\dagger} \hat{a}_{1}}{\hat{a}_{1}^{\dagger} \hat{a}_{1}+
\hat{a}_{2}^{\dagger} \hat{a}_{2}}\Bigr)^{n_5}\,
\Bigl(1+i\frac{\hat{a}_{1}^{\dagger} \hat{a}_{2}-
\hat{a}_{2}^{\dagger} \hat{a}_{1}}{\hat{a}_{1}^{\dagger} \hat{a}_{1}+
\hat{a}_{2}^{\dagger} \hat{a}_{2}}\Bigr)^{n_6}\, : \vert M\rangle~~~.
\end{array}
\label{fock.5}
\end{equation}
where in Eq.\,(\ref{fock.4})
\begin{equation}
\begin{array}{rl}
\mu_{(n_1+r)}^{(n_2+p)}=&\langle M\vert :
(\hat{a}_{1}^{\dagger}\hat{a}_{1})^{n_1+r}\,
(\hat{a}_{2}^{\dagger}\hat{a}_{2})^{n_2+p}:\vert M\rangle \\
\\
&=
\sum_{m=0}^{M}\,\frac{M!}{(m-n_1-r)!\,(M-m-n_2-p)!}\,
(\cos{\gamma})^{2m}\,(\sin{\gamma})^{2(M-m)}
\end{array}
\label{fock.6}
\end{equation}
with $n_{1}+n_{2} \le M$ in Eq.\,(\ref{fock.4}) 
and $\tilde{n} \le M$ where $\tilde{n}=n_{3}+n_{4}+n_{5}+n_{6}$ 
in Eq.\,(\ref{fock.5}). In Eq.\,(\ref{fock.5}), 
$\langle \hat{\cal E}_{\phi}(x,\phi_0) \rangle$ is understood in the same 
sense as $\langle \hat{\cal E}_{\phi}(x,\delta_0) \rangle$ in 
Eq.\,(\ref{meas.13}). 

The normalizations are 
determined as before by satisfying the condition 
$\langle \hat{\cal E}_{\theta}(0)\rangle=
\langle \hat{\cal E}_{\phi}(0,\phi_0)\rangle=1$. 

The simplest analytic results can be obtained for the case $M=1$ with   
$\gamma$ and $\phi_0$ being free parameters. This corresponds for the 
initial state to 
\begin{equation}
\vert 1 \rangle_{\phi_0,\gamma}=
\cos{\gamma}\,\vert 1 \,,\, 0 \rangle+e^{i\phi_{0}}\,\sin{\gamma}\,
\vert 0 \, ,\, 1 \rangle~ 
\label{fock.8}
\end{equation}
which is a fully polarized version of the split photon state in 
[2,6].  
Using Eqs\,(\ref{fock.4}-\ref{fock.6}) we find for the moments 
\begin{equation}
\langle \hat{\cal E}_{\theta}(x)\rangle=
\frac{{\cal N}_{\theta}}{2}\,[\cos^{2}{\gamma}+(-)^{x}\sin^{2}{\gamma}]~, 
\label{fock.9}
\end{equation}
where ${\cal N}_{\theta}^{-1}=1/2$ and 
\begin{equation}
\langle \hat{\cal E}_{\phi}(x),\phi_0\rangle=
\frac{{\cal N}_{\phi}}{8}\,\Bigl\{\,
[1+(-)^{x}+i^{x}+(-i)^{x}]+[1-(-)^{x}]\,\sin{2\gamma}\,\cos\phi_{0}+
[i^{x}-(-i)^{x}]\,\sin{2\gamma}\,\sin{\phi_0}
\Bigr\}~,
\label{fock.10}
\end{equation}
where ${\cal N}_{\phi}^{-1}=1/2$. 
For the probability distributions $P(\theta)$ and $P(\phi,\phi_0)$  
we use Eqs.\,(\ref{meas.14}-\ref{meas.16.b}) in the same spirit as we   
applied to the coherent initial state in Sec.\,II\,A. A  
simple calculation yields that 
\begin{equation}
P(\theta)=
\cos^{2}{\gamma}\,\delta(\theta)+
\sin^{2}{\gamma}\,\delta(\theta-\pi)
\label{fock.11}
\end{equation} 
and
\begin{equation}
P(\phi,\phi_0)=\frac{1}{2\pi}\,\Bigl\{1+\sin{2\gamma}\,
\cos(\phi_{0}-\phi)\Bigr\}
\label{fock.12}
\end{equation}
where the probability distributions are positive definite and properly 
normalized, i.e. 
$\int_{0}^{\pi}\,d\theta\,P(\theta)=
\int_{-\pi}^{\pi}\,d\phi\,P(\phi,\phi_0)=1$. 
At this point, a crucial limiting case in Eq.\,(\ref{fock.11}) and 
(\ref{fock.12}) needs to be mentioned. For $\gamma=\pi/4$, 
Eqs\,(\ref{fock.11}) and (\ref{fock.12}) describe the probability 
distributions of a fully polarized symmetric Fock state. For 
$\gamma=0$ and $\pi/2$ we have single mode photon Fock states 
$\vert 1,0\rangle$ and $\vert 0,1\rangle$. The measured probability 
distributions in these states are 
$P(\theta)=\delta(\theta)$, 
$P(\phi,\phi_0)=1/2\pi$ for $\gamma=0$ and, 
$P(\theta)=\delta(\theta-\pi)$,   
$P(\phi,\phi_0)=1/2\pi$ for $\gamma=\pi/2$; which, correctly 
describe the statistics of the single mode Fock state consistently with  
the theoretical expectations of a uniform distribution for 
$P(\phi,\phi_0)$. For all other $\gamma$, Eqs\,(\ref{fock.11}) and 
(\ref{fock.12}) correctly describe the theoretical distributions for a 
general $\vert M\rangle_{\phi_0,\gamma}$. 
This behaviour of the probability distributions 
can also be observed in Eq.\,(\ref{coh.8}) in the limits $1 \ll \eta$ 
and $\eta \ll 1$. The analytic 
calculations become exponentially harder for $2 \le M$. Nevertheless, 
explicit forms of the $P(\theta)$ and $P(\phi,\phi_0)$ 
can be given for a general $M$ as 
\begin{equation}
P(\theta)={\cal N}_{\theta}\,
\sum_{n_{1},n_{2}}^{\prime}\,\delta(\theta-\theta_{\{n\}})\,
\frac{1}{2^{n_1+n_2}\,n_1!\,n_2!}\,
\Bigl\{\sum_{r,p}\,\frac{(-1)^{r+p}}{2^{r+p}\,r!\,p!}\,
\mu_{(n_1+r)}^{(n_2+p)}\,\Bigr\}
\label{fock.13}
\end{equation}
where Eq.\,(\ref{fock.6}) is used, and 
\begin{equation}
\begin{array}{rl}
P(\phi,\phi_0)={\cal N}_{\phi}\,&\sum_{\{n\}}^{\prime}\,
\frac{1}{8^{n_3+n_4+n_5+n_6}\,n_3!\,n_4!\,n_5!\,n_6!} \,
\sum_{r=0}^{\infty}(\frac{-1}{2})^{r}\frac{1}{r!}\sum_{p=0}^{r}
{r \choose p} \\
&\times \sum_{\ell_3=0}^{n_3}{n_3 \choose \ell_3}
\sum_{\ell_4=0}^{n_4}{n_4 \choose \ell_4}
\sum_{\ell_5=0}^{n_5}{n_5 \choose \ell_5}
\sum_{\ell_6=0}^{n_6}{n_6 \choose \ell_6} \\
&\times \sum_{k_3=0}^{n_3}{n_3 \choose k_3}
\sum_{k_4=0}^{n_4}{n_4 \choose k_4}
\sum_{k_5=0}^{n_5}{n_5 \choose k_5}
\sum_{k_6=0}^{n_3}{n_3 \choose k_6} (-1)^{\ell_4+k_4}\,
(i)^{\ell_6+k_5-\ell_5-k_6}\\
&\times \sum_{m=0}^{M}\,e^{-i(\phi_{0}-\phi+\phi_{\{n\}})\,
(\tilde{\ell}-\tilde{k})}\,
\frac{M!}{(m-\tilde{\ell}-r+p)!~(M-m-p-\tilde{n}+\tilde{\ell})!}\,
(\cos{\gamma})^{2m-\tilde{\ell}+\tilde{k}}\,(\sin{\gamma})^{2(M-m)+
\tilde{\ell}-\tilde{k}}
\end{array}
\label{fock.14}
\end{equation}
with $\theta_{\{n\}}$ and $\phi_{\{n\}}$ as given 
by Eqs.\,(\ref{meas.6}). The numerical calculations of Eqs.\,(\ref{fock.13}) 
and (\ref{fock.14}) for linear polarization (e.g. $\phi_{0}=0$), for  
$\eta=1.0,0.5$ [i.e., corresponding to $\gamma=\pi/4, \tan^{-1}(\sqrt{2})]$ 
are presented in Figs.\,9 and 10 for various values of M. 
Like in the coherent case,  
the temporal phase factor $\phi_{0}$ in Eq\,(\ref{fock.14}) only shifts the 
distribution and does not play any role in the fluctuations. 
We now shift our attention to the second order fluctuations in the 
$\theta$ and $\phi$ dependent moments.

\subsubsection{The measured fluctuations in polarization}
Similar to the coherent state example in Sec. II A, 
we can examine the $\theta$ and $\phi$ dispersions in the weak-field  
limit in the range $0 \le \theta \le \pi$ and 
$-\pi \le \phi \le \pi$ using the same observables as in Sec.II.A.1.
For $M=1$ we have for $\theta$ 
\begin{equation}
\langle \hat{C}_{\theta}\rangle=\cos{2\gamma}~,\qquad 
\langle \hat{C}_{\theta}^{2} \rangle=1
\label{fock.15}
\end{equation}
hence
\begin{equation}
D_{\theta}=\sqrt{\langle (\hat{C}_{\theta}-\langle
\hat{C}_{\theta} \rangle)^{2}\rangle}=\sin{2\gamma}=
\frac{2\eta^{1/2}}{1+\eta}
\label{fock.15.b}
\end{equation}
and for $\phi$
\begin{equation}
\langle \hat{C}_{\phi}\rangle=
\frac{1}{2}\,\sin{2\gamma}\,\cos{\phi_{0}}~,\qquad 
\langle \hat{S}_{\phi}\rangle=
\frac{1}{2}\,\sin{2\gamma}\,\sin{\phi_{0}}~,\qquad 
\langle \hat{C}_{\phi}^{2}\rangle=
\langle \hat{S}_{\phi}^{2}\rangle=\frac{1}{2}
\label{fock.16}
\end{equation}
hence
\begin{equation}
D(\phi)=\sqrt{\langle (\hat{C}_{\phi}-\langle
\hat{C}_{\phi} \rangle)^{2}\rangle+
\langle (\hat{S}_{\phi}-\langle
\hat{S}_{\phi} \rangle)^{2}\rangle}=\sqrt{1-\frac{1}{4}\sin^{2}{2\gamma}}=
\sqrt{1-\frac{\eta}{(\eta+1)^{2}}} \ge \frac{\sqrt{3}}{2}~.  
\label{fock.16.b}
\end{equation}
It is not an accident that the weak-field limit
for the coherent state described in 
(\ref{cohfl.4}) coincides with the Fock state calculations in
(\ref{fock.16.b}) for $M=1$ for all $\eta$.
The results of the numerical calculations of $D(\theta)$ and $D(\phi)$  
as the initial number of photons is varied are shown for $\eta=1.0,0.1,0.01$   
in Fig.\,11 for $D(\theta)$ and, Fig.\,12 for $D(\phi)$ 
corresponding to linear polarization (e.g. $\phi_{0}=0$).  
Due to the large number of summations in Eq.\,(\ref{fock.14}) calculations 
are considered within the range $1 \le M \le 10$.   
 
The fully polarized Fock state is a typical example where the correlations  
are present between the relative occupations of the polarization components. 
As pointed out in Ref.\,[2],  
this renders the physical interpretation of the theoretically 
inferred moments for the 
$\phi$-related operators impossible. It appears that the operational 
approach here provides a scheme where the temporal phase distribution can  
be measured even if such correlations are present. We believe that the 
results obtained in Figs.\,9 and 10 should be checked experimentally   
with the particular emphasis on the weak-field regimes, which we expect to 
provide further confirmation of NFM's operational scheme. The 
split-photon state 
discussed by NFM in [2,6] can also be interpreted as the weak-field 
limit of the polarized Fock state in Eq.\,(\ref{fock.1}) corresponding to 
$M=1$ where strong intensity correlations are present. For this state  
the inferred moments of the corresponding $\hat{C}_{\phi}$ and 
$\hat{S}_{\phi}$ 
are unphysical because of the fact that in Eq.\,(\ref{meas.18}) the 
denominator vanishes. To examine the theoretically inferred moments for a 
general $M$ 
we use Eqs\,(\ref{meas.17}) and (\ref{meas.18}) in the Fock state 
(\ref{fock.1}) (we consider $\gamma=\pi/4$ for simplicity) to calculate  
\begin{equation}
\langle {(\hat{C}_{\theta}^{I})}^{x}\rangle=
\frac{\langle M \vert :(\hat{n}_{1}-\hat{n}_{2})^{x}:\vert M \rangle}
{\langle M \vert :(\hat{n}_{1}+\hat{n}_{2})^{x}:\vert M \rangle}
\label{fock.18}
\end{equation}
where we find that
\begin{equation}
\langle \hat{C}_{\theta}^{I}\rangle=
\langle {(\hat{C}_{\theta}^{I})}^{2}\rangle=0
\label{fock.19}
\end{equation}
and
\begin{equation}
\langle \hat{\cal E}^{I}_{\phi}(x)\rangle=
\frac{\langle M \vert : \Bigl[
(\hat{n}_{3}-\hat{n}_{4})+i\,(\hat{n}_{5}-\hat{n}_{6})\Bigr]^{x}:
\vert M\rangle}
{\langle M\vert : \Bigl[
\sqrt{(\hat{n}_{3}-\hat{n}_{4})^{2}+
(\hat{n}_{5}-\hat{n}_{6})^{2}} \Bigr]^{x}:\vert M \rangle}=
\frac{\langle :\Bigl[ (\hat{a}_{1}^{\dagger}\hat{a}_{2}+
\hat{a}_{2}^{\dagger}\hat{a}_{1})+i(\hat{a}_{1}^{\dagger}\hat{a}_{2}-
\hat{a}_{2}^{\dagger}\hat{a}_{1})\Bigr]^{x}:\rangle}
{\langle : \Bigl[\sqrt{\hat{a}_{1}^{\dagger}\hat{a}_{2}^{\dagger}
\hat{a}_{1}\hat{a}_{2}} \Bigr]^{x}:\rangle}~. 
\label{fock.20}
\end{equation}
The vacuum fields do not contribute to the normal ordering and 
we also omitted the state label $M$ in the second step of the expression. 
In order to calculate Eq.\,(\ref{fock.20}) we need 
$\langle: \Bigl[\sqrt{\hat{n}_{1}\hat{n}_{2}}\Bigr]^{x}:\rangle$.  
In the presence of correlations (i.e. $\langle\hat{n}_{1}
\hat{n}_{2}\rangle \ne \langle \hat{n}_{1}\rangle \langle \hat{n}_{2}
\rangle$) we have 
\begin{equation}
\frac{\langle \hat{n}_{1} \hat{n}_{2} \rangle}
{\langle \hat{n}_{1}\rangle \langle \hat{n}_{2} \rangle} 
=1-\frac{1}{M} < 1~. 
\label{fock.21}
\end{equation}
Hence, the correlation effects cannot be ignored if the  
initial Fock state contains a few number of photons. Furthermore, 
Eq.\,(\ref{fock.21}) implies that, for $M=1$ 
the denominator in Eq.\,(\ref{fock.20}) diverges at $x=2$. 
It might therefore be suggested to consider that the comparison with the 
theoretically inferred moments 
with the measured operational ones is limited to the strong-field   
regime ($1 \ll M$) where also consistency with the classical results are 
expected to hold. On the other hand, the denominator 
in Eq.\,(\ref{fock.20}) is not well-defined for values of $x$ not equal to 
an even integer. Now let us assume for the moment that we are able to 
replace the denominator of Eq.\,(\ref{fock.20}) by 
$(\langle \hat{a}_{1}^{\dagger}\hat{a}_{2}^{\dagger}
\hat{a}_{1}\hat{a}_{2}\rangle)^{x/2}$. One expects that if this 
replacement can be done, it can only be valid 
in the sufficiently strong-field limit where the correlations 
as well as fluctuations are expected  to be negligible. With this 
replacement, Eq.\,(\ref{fock.20}) would yield 
\begin{equation}
\langle \hat{C}_{\phi}^{I}\rangle=\frac{\cos{\phi_{0}}}{\sqrt{1-1/M}}
~,\qquad 
\langle \hat{S}_{\phi}^{I}\rangle=\frac{\sin{\phi_{0}}}{\sqrt{1-1/M}}~, 
\qquad 
\langle {(\hat{C}_{\phi}^{I})}^{2}\rangle=\cos^{2}{\phi_{0}}~,\qquad
\langle {(\hat{S}_{\phi}^{I})}^{2}\rangle=\sin^{2}{\phi_{0}}
\label{fock.22}
\end{equation}
The inferred dispersion $D(\phi)^{I}$ calculated from Eqs\,(\ref{fock.22}) 
is purely imaginary {\it for~all} $M$ which is an unphysical result. 
Hence, the replacement we made above, in order to make the denominator of 
Eq.\,(\ref{fock.20}) calculable, is unphysical for all $M$; thus, it cannot  
be done. Unlike the coherent state, the comparison with the theoretically 
inferred moments is made impossible by the presence of strong  
correlations. Therefore we are unable to examine the 
photodetector effects in the weak-field limit in the operational measurement 
of the fully polarized Fock state using the standard formalism of 
theoretically inferred moments. The unphysical results we obtained for 
the inferred moments are not inherent to the quantum scheme. Even in the 
classical measurement scheme, there is no unique way of extracting the 
theoretically inferred moments when the relative phase or the relative 
intensity fluctuations are correlated. We refer the reader to Ref.\,[2] 
for a detailed discussion on this topic in the context of operational 
phase formalism.  
Nevertheless, we will suggest in the following subsection that for the Fock   
state in Eq.\,(\ref{fock1.c}), or specifically for Eq.\,(\ref{fock.1}),  
it is possible to find another measure to examine the 
photodetector effects in the weak-field limit by making use of the 
properties of the uncertainty relations.   

\subsection{The fully polarized Fock state and connections to the $su(2)$  
interferometry}
For symmetric distribution of photon numbers in the components of the 
polarization, the fully polarized Fock state in Eq.\,(\ref{fock1.c}) 
becomes Eq.\,(\ref{fock.1}) which  
is a generalized $su(2)$ coherent state\cite{20}
\begin{equation}
\vert j_0 \,\xi\rangle=e^{(\xi \hat{J}_{+}-\xi^{*} \hat{J}_{-})}\,
\vert j_0 \,,\,-j_0\rangle=\frac{1}{(1+\vert \xi\vert^{2})^{j_0}}\,
\sum_{n=-j_0}^{j_0}\,
{2j_0 \choose j_0+n}^{1/2}\,\xi^{j_0+n}\,\vert j_0 \, n\rangle
\label{su2.1}
\end{equation}
which becomes clear if one makes a correspondence between 
Eqs\,(\ref{su2.1}) and (\ref{fock.1}) as 
\begin{equation}
\begin{array}{rl}
j_0 \to M/2~, \qquad
&j_0-n \to m~, \qquad  
j_0+n \to (M-m)~, \qquad
\xi \to e^{i\phi_{0}} ~,\qquad {\rm or} \\
&j_0+n \to m~, \qquad 
j_0-n \to (M-m)~, \qquad
\xi \to e^{-i\phi_{0}}
\end{array}
\label{su2.2}
\end{equation}
where in Eq.\,(\ref{su2.1}) $\vert j_0\,n\rangle=
\vert j_0-n\,,\, j_0+n \rangle$. Here, $\hat{J}_{\pm}$ are the standard  
$su(2)$ raising and lowering operators of the $su(2)$ angular momentum 
algebra defined by the generators $\hat{J}_{i}~,(i=0,\dots,3)$
\begin{equation}
\begin{array}{rl}
\hat{J}_{0}=&(\hat{a}_{1}^{\dagger}\,\hat{a}_{1}+
\hat{a}_{2}^{\dagger}\,\hat{a}_{2})/2 \\
\hat{J}_{1}=&(\hat{a}_{1}^{\dagger}\,\hat{a}_{1}-
\hat{a}_{2}^{\dagger}\,\hat{a}_{2})/2 \\
\hat{J}_{2}=&(\hat{a}_{1}^{\dagger}\,\hat{a}_{2}+
\hat{a}_{2}^{\dagger}\,\hat{a}_{1})/2 \\
\hat{J}_{3}=&(\hat{a}_{1}^{\dagger}\,\hat{a}_{2}-
\hat{a}_{2}^{\dagger}\,\hat{a}_{1})/2i~, 
\end{array}
\label{su2.3}
\end{equation}
where, considering that $(\hat{a}_{1}^{\dagger},\hat{a}_{1}),
(\hat{a}_{2}^{\dagger},\hat{a}_{2})$ represent two independent boson pairs,
we have the standard $su(2)$ algebra
$[\hat{J}_{i},\hat{J}_{j}]=i\epsilon_{ijk}\,\hat{J}_{k}~~,(i,j,k=1,2,3)$.
Here the central invariant of the algebra is 
$\hat{J}^{2}=\hat{J}_{0}(\hat{J}_{0}+1)=j_{0}(j_{0}+1)$ where $j_{0}=M/2$.
Here, $M$ being an eigenvalue of $\hat{J}_{0}$
describes the total number of particles in the Fock state (\ref{fock.1}).
The uncertainty relations for $\hat{J}_{i}$'s are given by 
\begin{equation}
(\Delta \hat{J}_{i})\,(\Delta \hat{J}_{j}) \ge
\frac{\vert \epsilon_{ijk}\vert}{2}\,\langle J_{k} \rangle~,\qquad
i \ne j \ne k=1,2,3~.
\label{su2.4}
\end{equation}
The fully polarized Fock state is nothing but the generalized 
coherent state of the free field $su(2)$ angular momentum algebra. 
Under certain conditions Eq\,(\ref{su2.1}) also coincides with the 
su(2) minimum uncertainty states\cite{21,22,23} minimizing 
Eq.(\ref{su2.4}); and, this fact has been recently   
explored in the current literature in the context of $su(2)$ 
interferometry.\cite{22,23,24}  

The idea of $su(2)$ interferometry is to create interference between two 
arbitrary input fields by using passive and active lossless optical devices 
to measure the relative temporal phase between the fields. For this purpose 
the measured operators of the $su(2)$ interferometry are defined as 
in Eqs\,(\ref{su2.3}) or they are related to Eqs\,(\ref{su2.3}) by 
certain unitary transformations induced by the passive and active optical 
devices. These transformations on Eqs\,(\ref{su2.3}) [or the inverse  
transformations on the initial fields] can be engineered 
in such a way that the relative phase shift between the input fields can be 
measured by pure intensity measurements on the fields at the output ports 
of the interferometer.\cite{22} The principles of the quantum 
interferometry are thus based on a generalized 
operational scheme that is, in principle, very similar to the idea 
of the operational phase measurement presented in [2,3,6,9] as well  
as the present work.   

Now let us construct the uncertainty product for the general fully polarized 
Fock state $\vert M\rangle_{\phi_0,\gamma}$ and particularly focus our 
attention on the specific limit $\vert M\rangle_{\phi_0}$ at 
$\gamma=\pi/4$. 
The measured interferometric operators correspond, in the standard $su(2)$ 
interferometry, to the expected values of the operators in Eqs(\ref{su2.3}) 
or some linear superpositions of them in the initial state. For 
$\vert M\rangle_{\phi_0,\gamma}$ being the initial state, we have 
\begin{equation}
\begin{array}{rl}
\langle \hat{J}_{0}\rangle=&\frac{M}{2} \\
\langle \hat{J}_{1}\rangle=&\frac{M}{2}\,\cos{2\gamma}  \\
\langle \hat{J}_{2}\rangle=&\frac{M}{2}\,\sin{2\gamma}\,\cos{\phi_0} \\
\langle \hat{J}_{3}\rangle=&\frac{M}{2}\,\sin{2\gamma}\,\sin{\phi_0} \\
\end{array}
\label{su2.5}
\end{equation}
and
\begin{equation}
\begin{array}{rl}
(\Delta \hat{J}_{0})^{2}=&0 \\
(\Delta \hat{J}_{1})^{2}=&\frac{M}{4}\,\sin^{2}2\gamma \\
(\Delta \hat{J}_{2})^{2}=&\frac{M}{4}\,(1-\cos^{2}\phi_0\,\sin^{2}2\gamma) \\
(\Delta \hat{J}_{3})^{2}=&\frac{M}{4}\,(1-\sin^{2}\phi_{0}\,\sin^{2}2\gamma)~. 
\end{array}
\label{su2.6}
\end{equation}
Using $\gamma=\pi/4$ in Eqs.\,(\ref{su2.6}) and (\ref{su2.5}) we observe 
that $\vert M\rangle_{\phi_0}$ 
is an important state in the algebra defined  
by the operators in Eq.\,(\ref{su2.3}). It is an $su(2)$ coherent state  
[see Eq.\,(\ref{su2.1})] as well as a minimum 
uncertainty (intelligent) state minimizing Eq.\,(\ref{su2.4})  
for $i\ne j=1,3; k=2$ and $i\ne j=1,2; k=3$. This can be explicitly seen   
by using Eqs.\,(\ref{su2.5}) and (\ref{su2.6}) in Eq.\,(\ref{su2.4}).  
Furthermore, when $\gamma=\pi/4$,
this result is independent from $\phi_{0}$; hence, 
a temporal shift in $\phi_0$ does not change any of these properties. 
This implies that, if $su(2)$-interferometric techniques\cite{22} 
are employed for $\vert M\rangle_{\phi_0}$, the standard precision  
can be achieved in the measurement of the temporal phase.\cite{23,24} The 
precision in the phase measurement can be found from Eqs\,(\ref{su2.5}) and 
(\ref{su2.6}) for the general case with $\vert M\rangle_{\phi_0,\gamma}$ as  
\begin{equation}
\delta \phi_{0}(\gamma)=\frac{(\Delta \hat{J}_{1})}
{\vert \partial \langle \hat{J}_{1}\rangle/\partial \phi_0\vert}=
\frac{1}{\sqrt{M}}\,\frac{\sqrt{1-\sin^{2}2\gamma\,\sin^{2}\phi_0}}
{\sin2\gamma\,\cos\phi_0}   
\label{su2.7}
\end{equation}
where $\delta(\gamma) \ge \delta\phi_0(\pi/4)=1/\sqrt{M}$ 
which is the well-known minimum standard noise limit. 
Hence, {\it theoretically}, the maximum precision in the phase measurement 
can be achieved only at $\gamma(\pi/4)=1$ corresponding to $\eta=1$. 
The basic idea   
being the extraction of the phase statistics from pure photon counting, 
the $su(2)$ interferometry is in a close analogy to the operational 
measurement scheme. The operators $\hat{J}_{i}~, ~~(i=0,1,2,3)$ are the 
interferometric analogs of the operational ones $\hat{\Sigma}_{i}~,~~
(i=0,1,2,3)$ in Eq.\,(\ref{I.7}), but 
there are also significant differences between them. Although the 
$\hat{J}_{i}$'s  
are the generators of the $su(2)$ algebra, the $\hat{\Sigma}_{i}$'s all 
commute with each other and no useful uncertainty product similar to 
Eq.\,(\ref{su2.4}) 
can be written for them. Now a legitimate question arises on how much the 
properties of the quantum state $\vert M\rangle_{\phi_0,\gamma}$ 
, as far as the $\hat{J}_{i}$'s 
are concerned, are preserved in the operational measurement scheme using 
the $\hat{\Sigma}_{i}$ operators. The main difference arising from the 
presence of the vacuum states in the $\hat{\Sigma}_{i}$'s as well as 
the operational scheme itself, it is nevertheless expected that 
for sufficiently strong fields the quantum operational measurement using 
the $\hat{\Sigma}_{i}$ operators should be consistent with 
Eq.\,(\ref{su2.4}). The deviations in the quantum operational measurement 
scheme from Eq.\,(\ref{su2.4}) are expected when the initial field is 
sufficiently weak. Hence, by   
examining the uncertainty properties of $\vert M\rangle_{\phi_0,\gamma}$,  
particularly near $\gamma=\pi/4$, a perfect ground to understand 
the influence of the operational scheme in the final measurement can 
be provided.   

We start the analysis of the uncertainty relations for 
$\vert M\rangle_{\phi_0,\gamma}$ by examining the $\gamma$ 
dependence of the measured $D(\theta)$ and $D(\phi)$. The results are 
represented in 
Fig.\,13 for $M=1,5$ and for linear polarization in the range $0 \le \gamma 
\le \pi/2$. The figure indicates that, similarly to the results obtained 
for the fully polarized coherent state 
measurements, it is not possible to simultaneously minimize the 
fluctuations in the measurements of the $\theta$ and $\phi$ related moments. 
The values of $D(\theta)$ and $D(\phi)$ in Fig.\,13 corresponding to 
$\gamma=\pi/4$ (i.e., $\eta=1$), $\gamma \simeq 0.3 \pi$ (i.e., $\eta=0.5$), 
$\gamma \simeq 0.4\pi$ (i.e., $\eta=0.1$), and 
$\gamma\simeq 0.47 \pi$ (i.e., $\eta=0.01$) can also be read from the 
Figs.\,11 and 12. Here $\gamma=\pi/4$ has a special importance since 
this point corresponds to where $\vert M\rangle_{\phi_0,\gamma}$ becomes  
a coherent as well as a minimum uncertainty state of the $\hat{J}_{i}$'s. 
As $\gamma$ is shifted away from $\pi/4$, particularly towards 
$\gamma=0, \pi/2$, one particular polarization mode starts dominating where 
the initial state gradually starts looking like a single mode 
Fock state. The single mode Fock limit is realized at $\gamma=0, \pi/2$ 
for $\theta$ 
related measurements [i.e., $D(\theta)=0$ and $P(\theta)$ comprises 
a single delta function peak], and a maximally random fluctuations are 
observed in the $\phi$ related measurements [i.e., $D(\phi)=1$ and 
$P(\phi)=1/2\pi$]. Because of the fact that 
the interferometric operators do not commute with each other, it is 
not possible to find the interferometric analogs of the trigonometric 
operators $\hat{C}_{\theta}, \hat{S}_{\theta}$ and 
$\hat{C}_{\phi}, \hat{S}_{\phi}$ defined in Eqs.\,(\ref{I.8}). This implies  
that, the interferometric analogs of $D(\theta)$ and $D(\phi)$ cannot be  
found by direct analogy and a comparison between the theory and the 
measurement is not possible for them. At this level, 
the only comparison with the theory can be made by examining the minimum 
uncertainty product for the $\hat{J}_{i}$'s and the $\hat{\Sigma}_{i}$'s. 

Keeping $\gamma$ as the parameter, we now express   
Eq.\,(\ref{su2.4}) in $\vert M\rangle_{\phi_0,\gamma}$ in the form 
\begin{equation}
{\cal U}(\gamma)=\frac{(\Delta \hat{J}_{2})^{2}+(\Delta \hat{J}_{3})^{2}}
{\langle \hat{J}_{2}\rangle^{2}+\langle \hat{J}_{3}\rangle^{2}}\,
(\Delta \hat{J}_{1})^{2}
\label{su2.8}
\end{equation}
and find from Eqs.\,(\ref{su2.5}) and (\ref{su2.6}) that
\begin{equation}
{\cal U}(\gamma)=\frac{1}{2}\,(1-\frac{1}{2}\,\sin^{2}2\gamma)
\label{su2.9}
\end{equation}
with the minimum uncertainty corresponding to ${\cal U}(\pi/4)=1/4$. The 
operational analog of Eq\,(\ref{su2.8}) in terms of the $\hat{\Sigma}_{i}$'s  
can be found by direct inspection of Eqs\,(\ref{I.5}-\ref{I.7})  
and (\ref{su2.3}) as
\begin{equation}
{\cal U}_{op}(\gamma)=\frac{(\Delta \hat{\Sigma}_{2})^{2}+
(\Delta \hat{\Sigma}_{3})^{2}}
{\langle \hat{\Sigma}_{2}\rangle^{2}+
\langle \hat{\Sigma}_{3}\rangle^{2}}\,(\Delta \hat{\Sigma}_{1})^{2}
\label{su2.10}
\end{equation}
with all fluctuations in Eq.\,(\ref{su2.10}) calculated within the 
operational scheme outlined in Sec.\,(II). In comparing 
Eq.\,(\ref{su2.10}) with Eq.\,(\ref{su2.8}) the differences arising from  
the different normalizing factors of the transmission and the reflection 
coefficients in the $\hat{d}_{i}$'s in Eq.\,(\ref{I.5}) should also be 
accounted for. The result of the numerical calculations for 
${\cal U}_{op}(\gamma)/
{\cal U}(\gamma)$ is presented in Fig.\,14 as a function of $M$. 
We also observed that Eq.\,(\ref{su2.10}) has no $\phi_{0}$ dependence 
for all $M$ and $\gamma$ 
[not shown in Fig.\,14] which is consistent with the 
theoretical calculation in Eq.\,(\ref{su2.9}). 

The Fig.\,14 indicates that the photodetection in the 
operational scheme unavoidably creates an additional noise in the 
measurement such that the theoretical value of the uncertainty is not 
reached until the initial state has sufficiently large 
(i.e. $5 \stackrel{<}{\sim} M$) number of photons. Here, as $\gamma$ 
varies, there is a 
compromise between the value of the measured uncertainty product for large 
$M$ 
and the detector noise for small $M$. For instance, at 
$\gamma=\pi/4$ (i.e., $\eta=1$), the measured uncertainty 
product approaches the theoretical minimum uncertainty for large $M$,    
although the detector {\it noise}  
is as large as $100 \%$ at the small $M$ limit. On the other hand, as 
$\gamma$ deviates from $\pi/4$, the measured uncertainty product is 
no more at the minimum for large $M$ but the detector noise is smaller for 
small $M$. 
Hence, it appears that there is no global optimum value for $\gamma$. We 
thus conclude that   
$\gamma$ can be optimally fixed only depending on the individual observables 
chosen in the measurement (i.e., a result which we have also reached in 
the fully polarized coherent state example in Sec. II.A.1). 

In the theoretical  
interferometric calculations it is a common practice to neglect the  
influence of the photodetection. This is certainly a valid assumption 
if the initial field is sufficiently strong. On the other   
hand, we expect the additional higher bound on the uncertainty product to  
be a manifestation of any scheme based on photon counting in the weak-field 
limit arising from the quantum nature of the photodetection. Hence it is 
also natural to expect these effects to be observable in the $su(2)$ 
interferometric measurements. This result indeed needs experimental 
verification, 
particularly, considering the advantage that certain schemes have been 
proposed for the generation of such quantum states as 
Eq.\,(\ref{fock.1}) 
experimentally using active non-linear processes.\cite{25}

\section{Discussion}
In this work, we focused our attention on the operational 
measurement scheme as applied to certain fully polarized quantum states  
particularly in the weak-field regime. We have shown that, similarly 
to NFM's operational phase measurement scheme 
it is possible to base the measurement of the state of 
polarization on pure photocount measurements, hence providing another 
example for operational approach. In particular, the measurement of the 
fluctuations 
of the temporal phase between the polarized field components is, not 
surprisingly, identical to the original work by NFM. 
The statistical behaviour of the Stokes parameters is investigated in 
terms of the trigonometric operators in Eqs.\,(\ref{I.8}) and the operational 
counterparts of the quantum Stokes parameters of the polarized field are 
introduced in Eqs.\,(\ref{I.7}). The application of the operational 
polarization measurement scheme is made to fully polarized quantum coherent 
as well as Fock states. With the purpose of extracting the detectors' 
influence on the measurement, the statistics of the measured 
fluctuations are examined and compared with the theoretical calculations. 
Our results confirm those of NFM's operational phase measurement scheme 
to conclude that the photon-counting process introduces additional noise 
in the final statistics particularly in the weak-field regimes. For 
sufficiently strong fields, the operational measurement scheme is 
consistent with those theoretical predictions in which the photo-detection  
effects are not included. 

The connection between the operational 
approach to the measurement of polarization and the $su(2)$ interferometry 
is examined and the uncertainty principle is used as a means of analyzing  
the photodetection effects in the measurement where applications are made 
on the fully polarized Fock state. 

Operational approach to the measurement of phase has been investigated 
by D'Ariano and Paris\cite{26} in the context of quantum estimation 
theory\cite{27} which provides a unified formulation of the measurement 
process and the initial system under investigation. The quantum  
probability distribution of the $N$-port homodyne detection in 
Eq.\,(\ref{meas.1}) is a specific example of the 
probability-operator-valued measure (POM) in the quantum estimation theory. 
An ideal quantum measurement is realized when the POM is based on an 
orthogonal and complete set of states comprising the eigenspace of the 
measured observable. Hence, depending on the nature of the measured 
observables of the initial state, finding an optimum detection scheme 
is the primary goal of a unified formulation of the measurement and the 
initial system. For the measured observable being phase related quantities,  
such an approach has not been idealized yet because of the fact that 
an orthogonal POM cannot be physically realized for the phase observable. 
With this in mind, one resorts to optimizing the phase measurement by 
a proper 
choice of the initial states as well as the parameters of the measuring 
system. At this point a connection is present between the primary goal of 
the quantum estimation theory and the attempts to surpass 
the standard noise limit by using interferometric transformations on the 
measuring system (or inverse transformations on the initial state).  
The optimal choice of these transformations
using active as well as passive optical devices naturally
depends on the initial state. Furthermore, it is also desirable (and under
certain conditions it is strictly required) to have 
the transformed state conserve the basic features of the original
state i.e., full polarization, coherent and minimum uncertainty state for
Eq.\,(\ref{fock.1}), the statistics of the fluctuations, etc.
It is natural that
for the initial state being fully polarized, the full polarization itself
is a strict condition that should be conserved by the transformations.
On the other hand, since $\vert M\rangle_{\phi_0}$ is a minimum
uncertainty state, the quantum statistics of the temporal phase $\phi$
and the fluctuations in corresponding $\phi$ dependent operators are
coupled with those describing the fluctuations in the relative photon number 
$\hat{n}_{1}-\hat{n}_{2}$. Hence, the minimum uncertainty condition of 
$\vert M\rangle_{\phi_0}$ will most certainly be at stake after such
transformations and this will change the quantum nature of the state
[for instance, a rotation in the field space by $\gamma$ does not change
the full polarization property but changes the
minimum uncertainty relations, as it can be seen from Eq.\,(\ref{su2.4})
and (\ref{su2.9})]. For those states which are not the minimum 
uncertainty ones, this observation is still valid to a lesser extent.
We nevertheless conclude that attempts to surpass the standard noise limit
for the fully polarized quantum states have to comply with a number of
restrictions, which certainly renders it to be rather interesting problem.

\newpage

\vskip5truecm
{\Large \bf Figure Captions:}
\\
\\
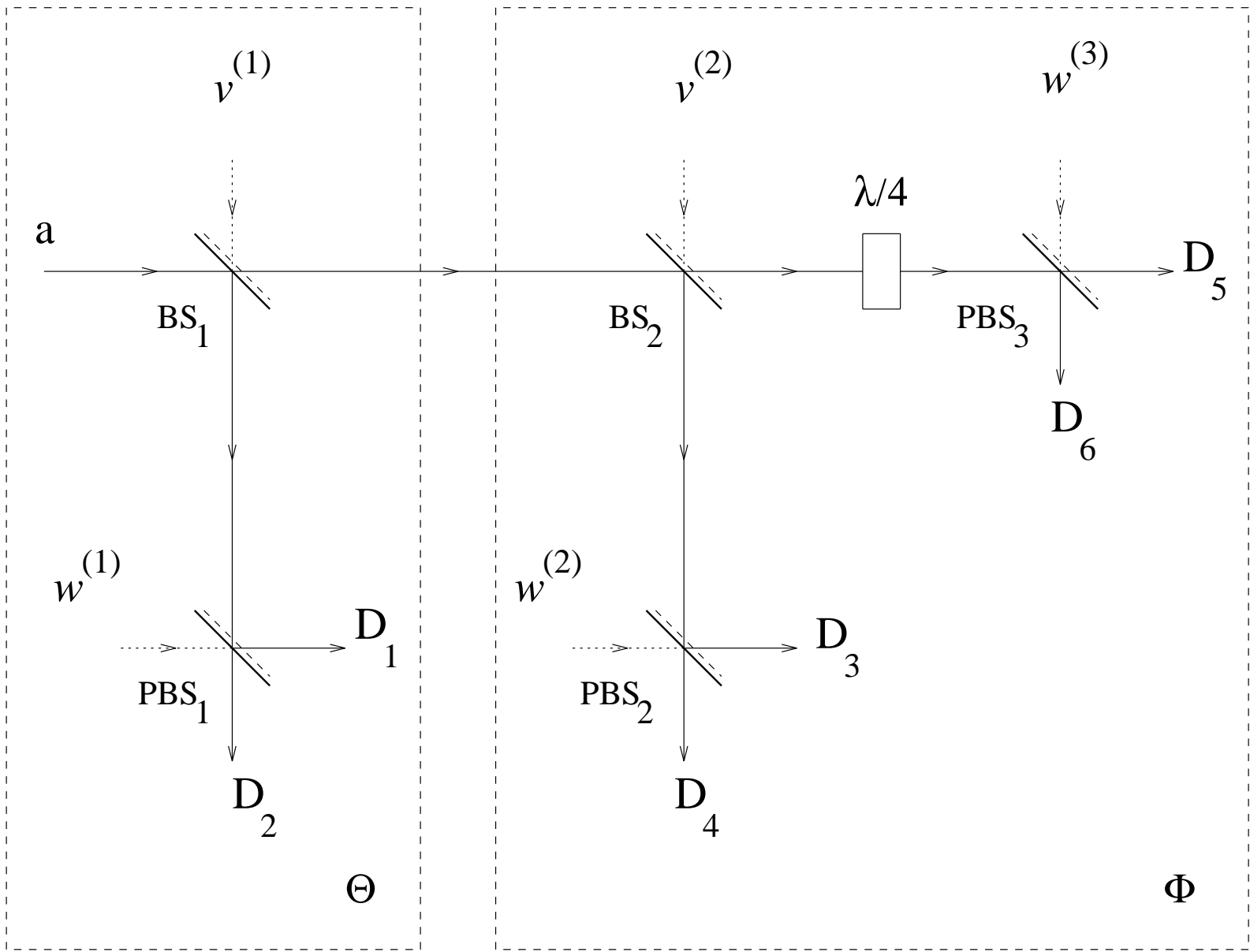
{\bf Fig.1} The experimental setup to measure the classical and 
quantum Stokes parameters. Note that this setup is also able to 
measure all of the total of six Stokes parameters for a partially 
polarized light. 
\\
\\
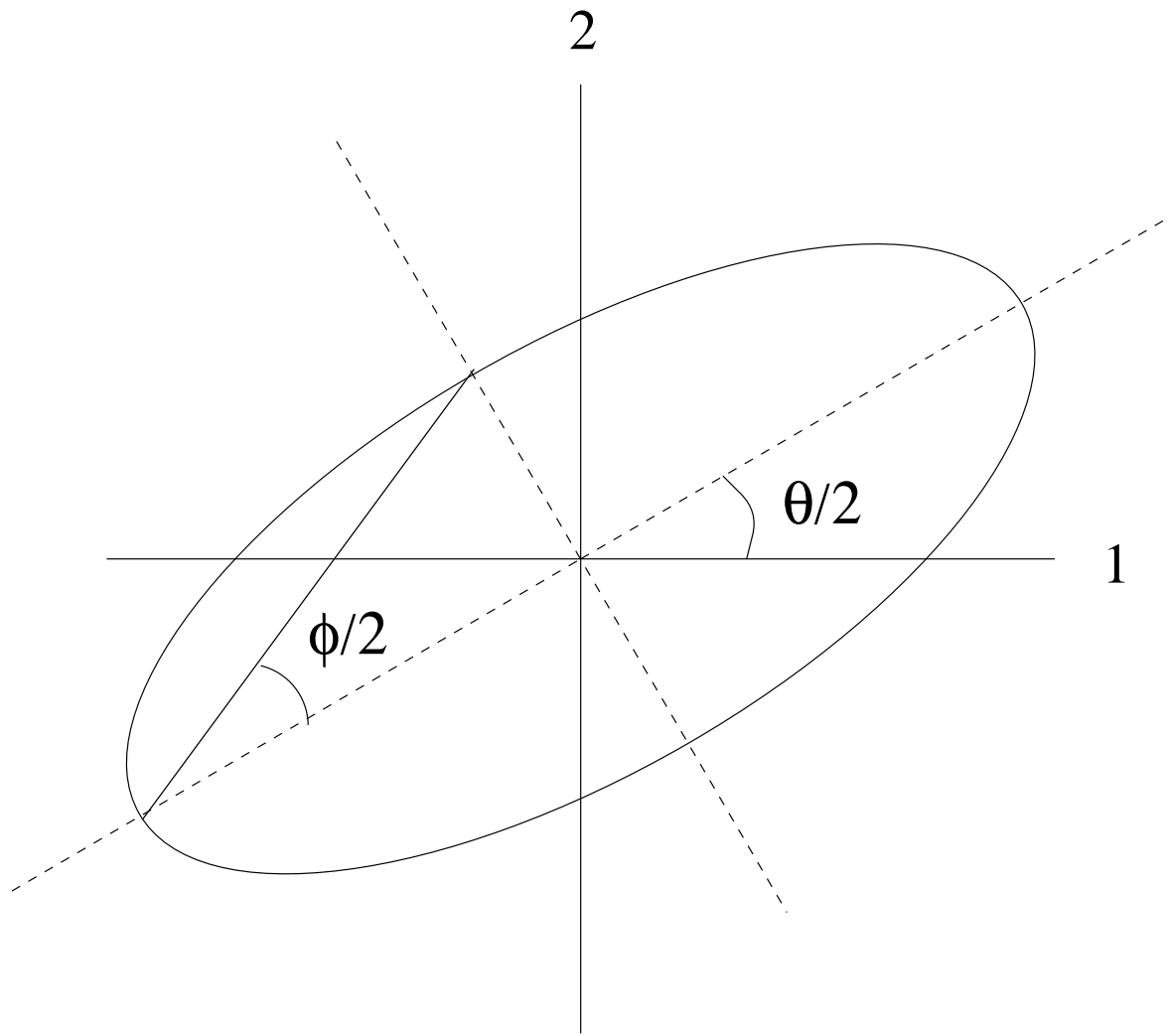
{\bf Fig.2} Ellipsometry for the fully polarized transverse electric  
field ${\bf E}$ in the tangent plane. Angular parameters are shown as 
defined in Eqs\,(\ref{I.4}).  
\\
\\
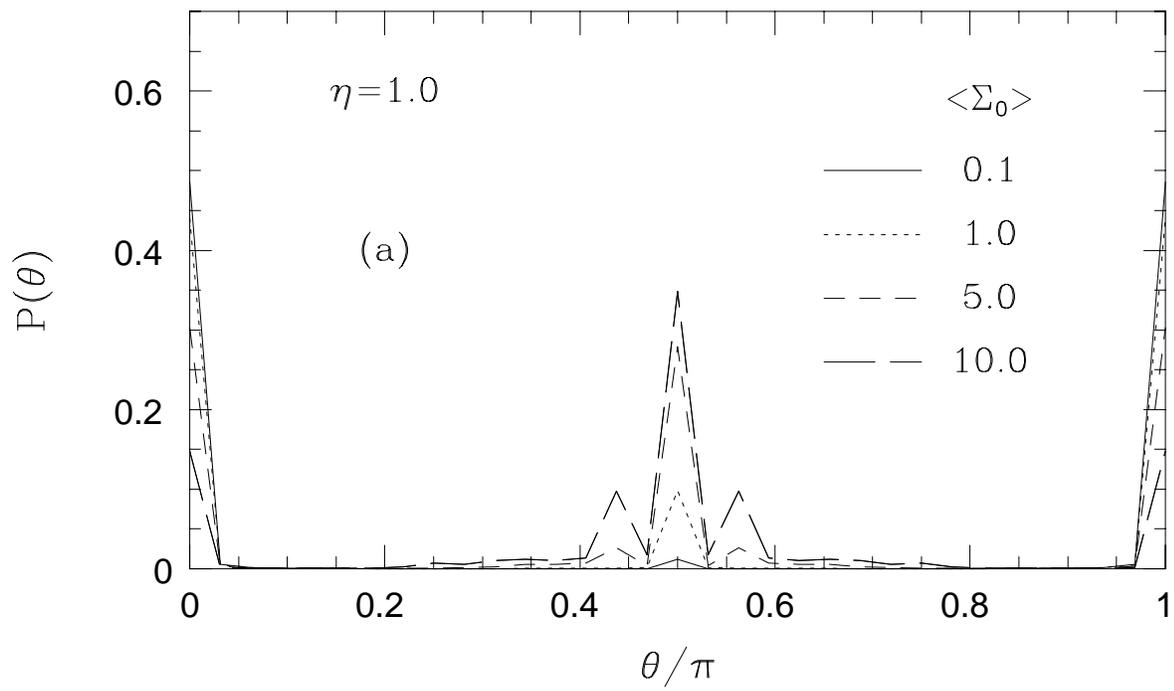
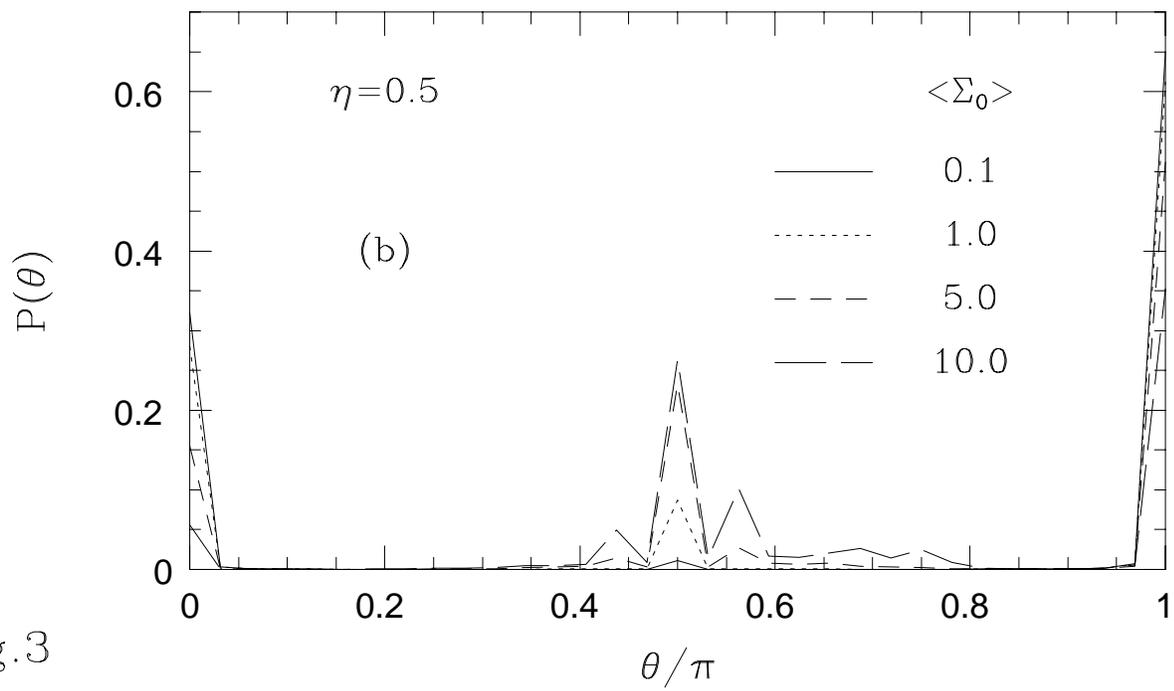
{\bf Fig.3} Measured probability distribution $P(\theta)$ 
versus $\theta$ for the fully polarized coherent state and for 
a) $\eta=1.0$, b) $\eta=0.5$ for the indicated average total photon numbers. 
\\
\\
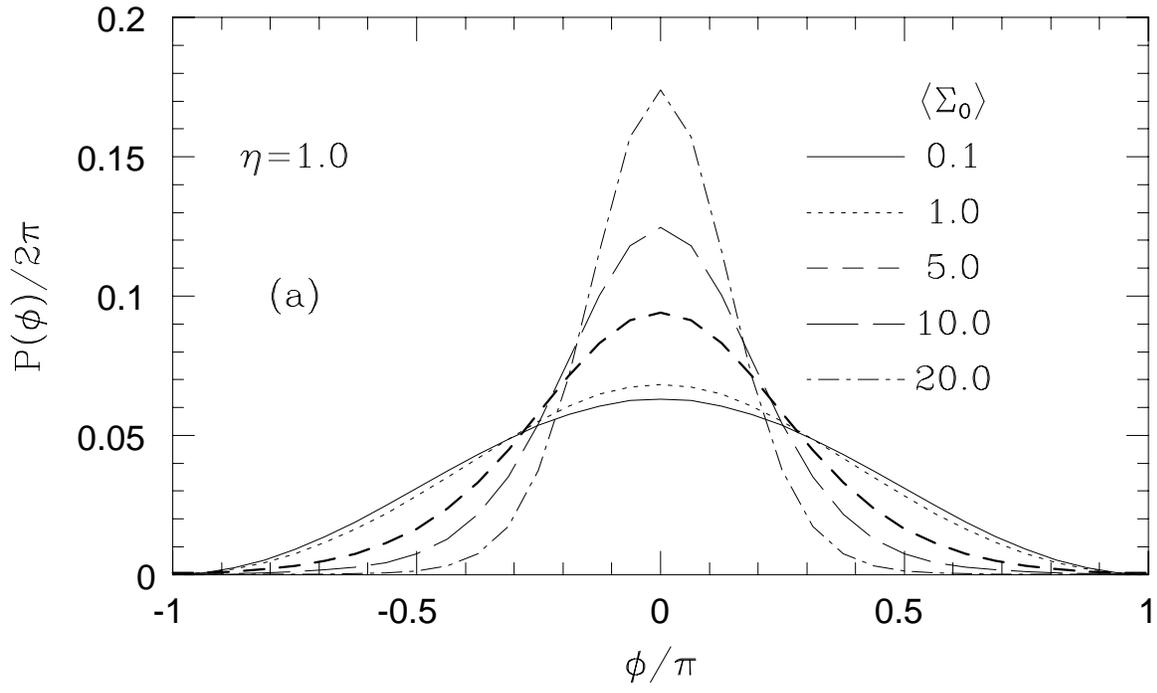
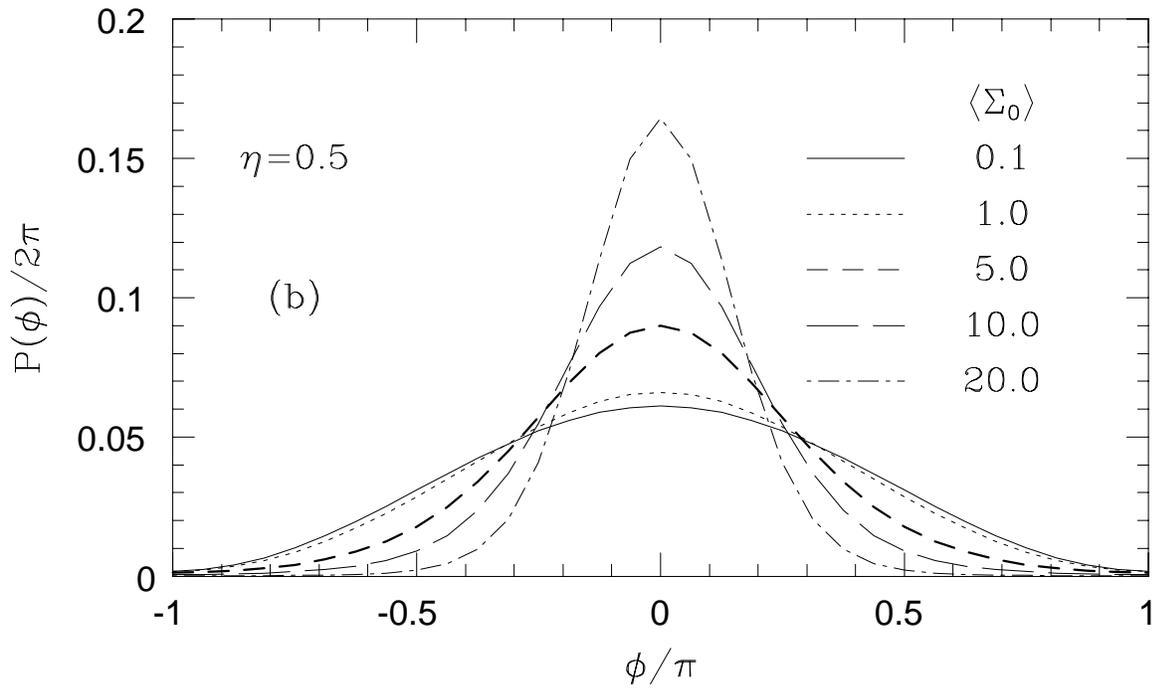
{\bf Fig.4} Measured probability distribution $P(\phi)$
versus $\phi$ for the fully polarized coherent state and for  
a) $\eta=1.0$, b) $\eta=0.5$ for the indicated average total 
photon numbers. 
\\
\\
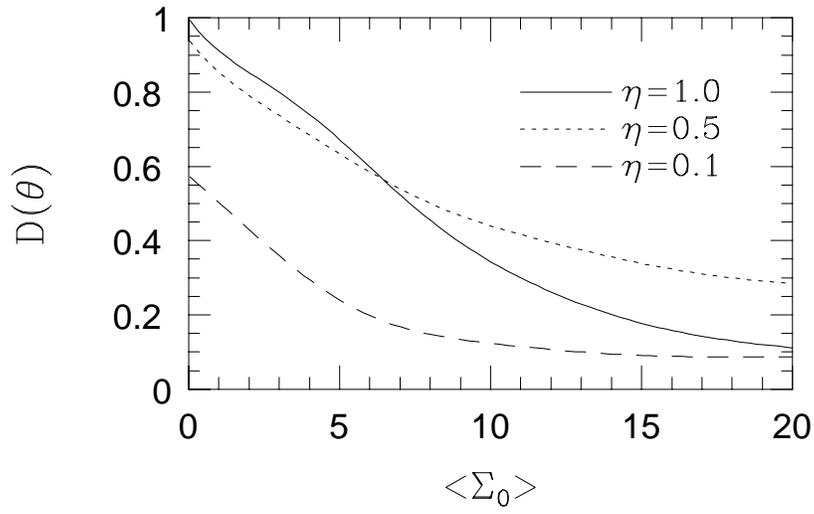
{\bf Fig.5} Second-order fluctuations in the $\theta$  
related measurement for the fully polarized coherent state and 
for the indicated values of $\eta$. 
\\
\\
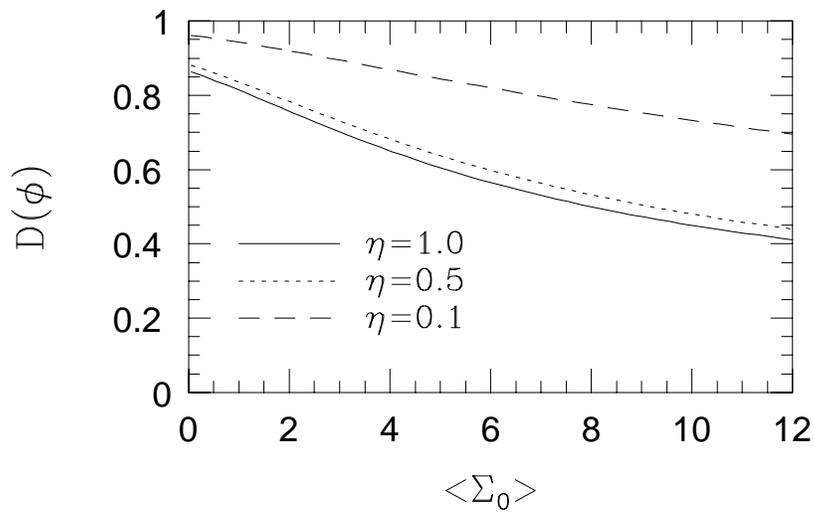
{\bf Fig.6} Second-order fluctuations in the $\phi$ 
related measurement for the fully polarized coherent state and 
for the indicated values of $\eta$.
\\
\\
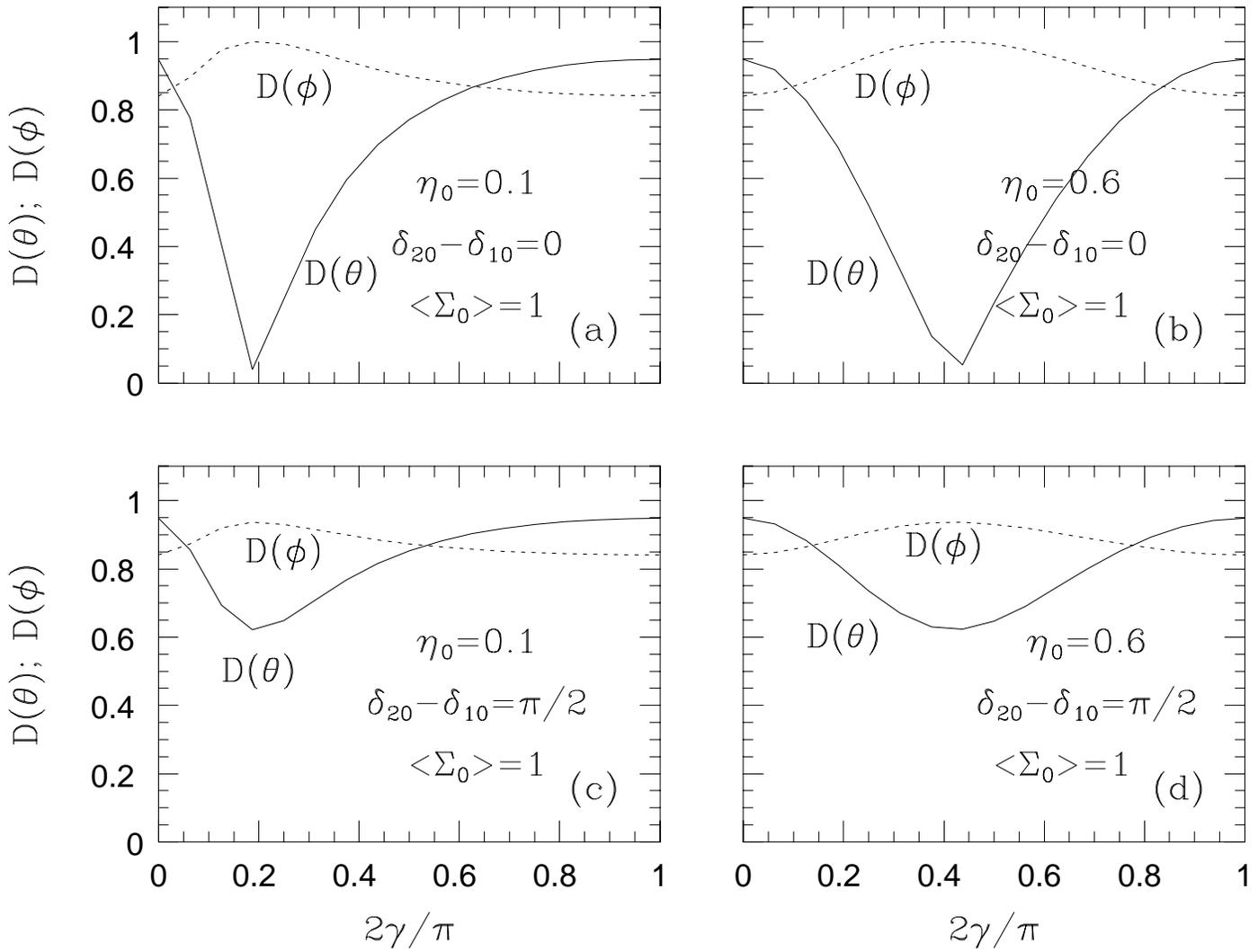
{\bf Fig.7} Second-order fluctuations in the $\theta$ and $\phi$ 
related measurements for the fully polarized coherent state 
as a function of the rotation parameter $\gamma$ 
for the indicated values of $\eta$ and the relative temporal phase 
in the extreme weak-field limit $\langle \Sigma_0\rangle=1$. 
\\
\\
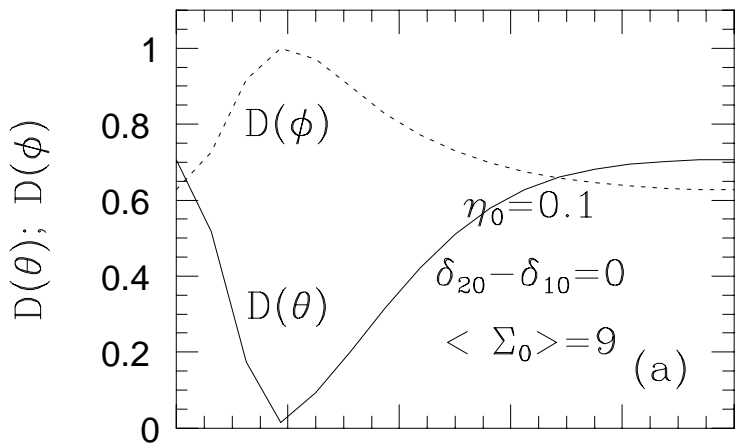
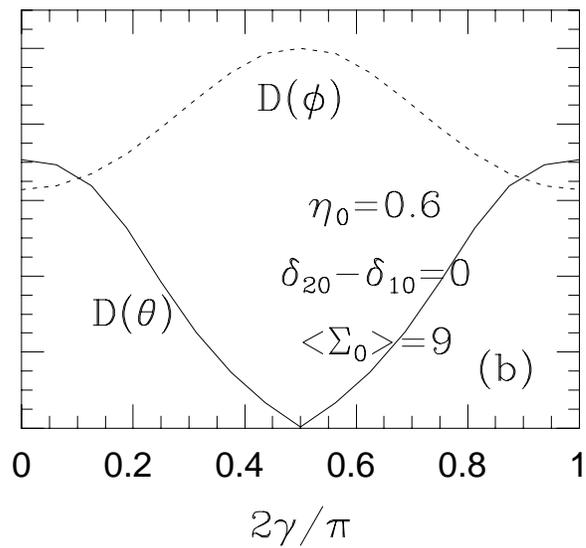
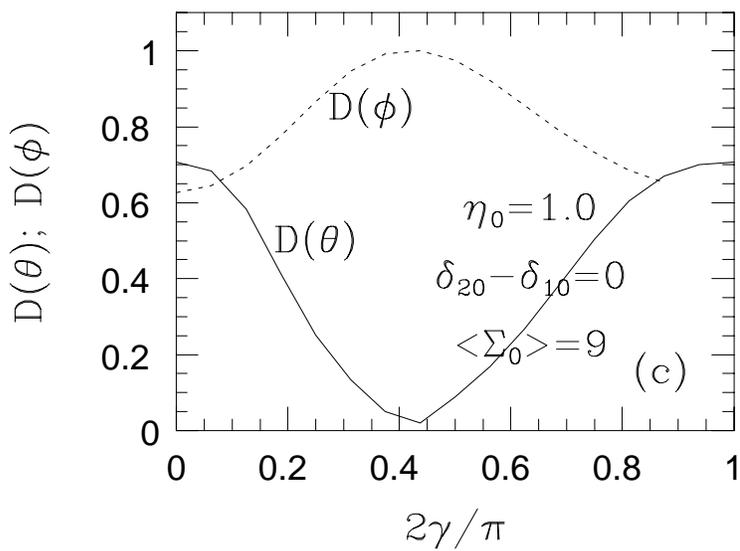
{\bf Fig.8} Second-order fluctuations in the $\theta$ and $\phi$ 
related measurements for the fully polarized coherent state 
as a function of the rotation parameter $\gamma$ 
for the indicated values of $\eta$ and for $\langle \Sigma_0\rangle=9$.
\\
\\
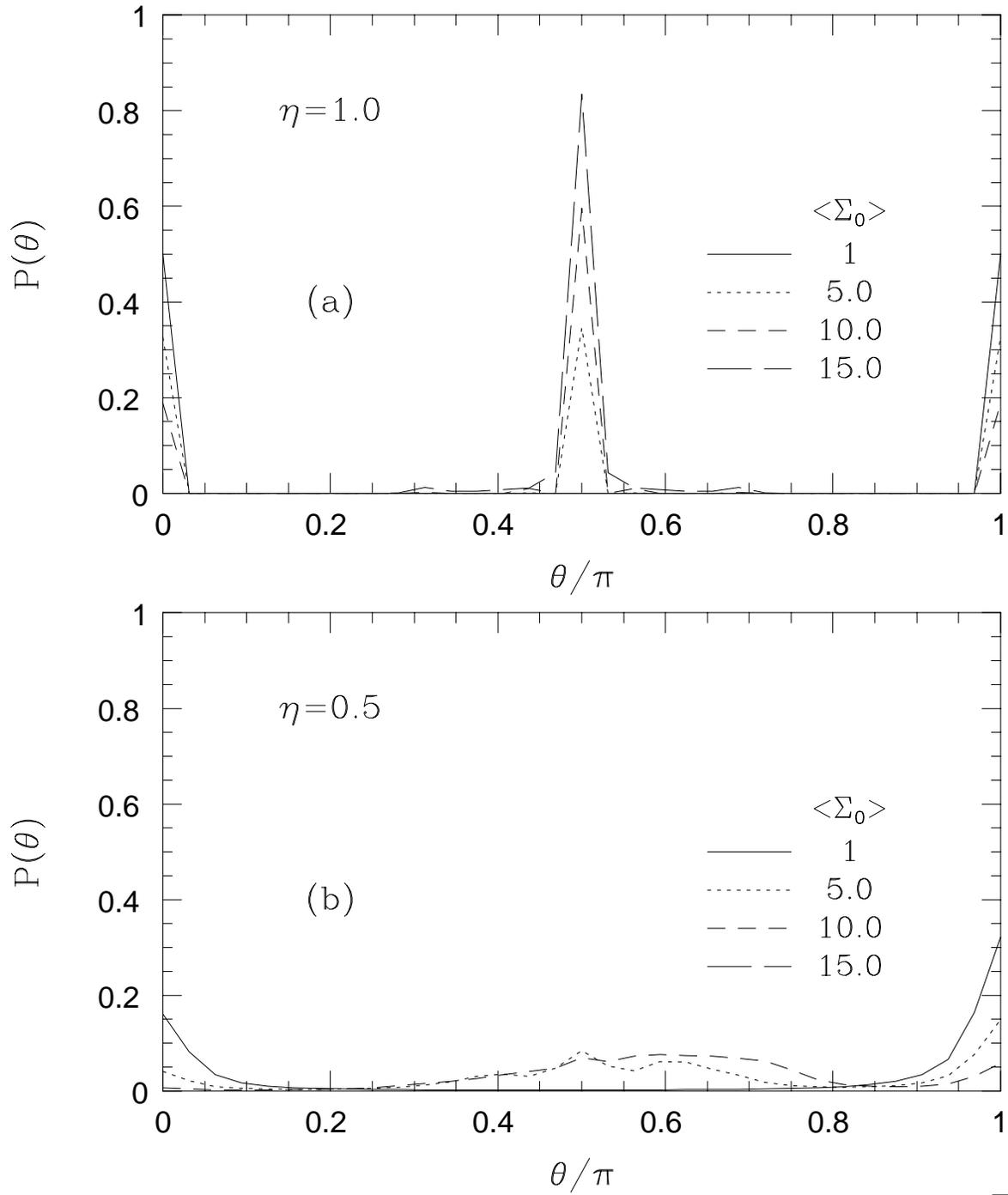
{\bf Fig.9} Measured probability distribution $P(\theta)$
versus $\theta$ for the fully polarized Fock state for 
a) $\eta=1$, b) $\eta=0.5$ and for the indicated average total photon numbers.  
\\
\\
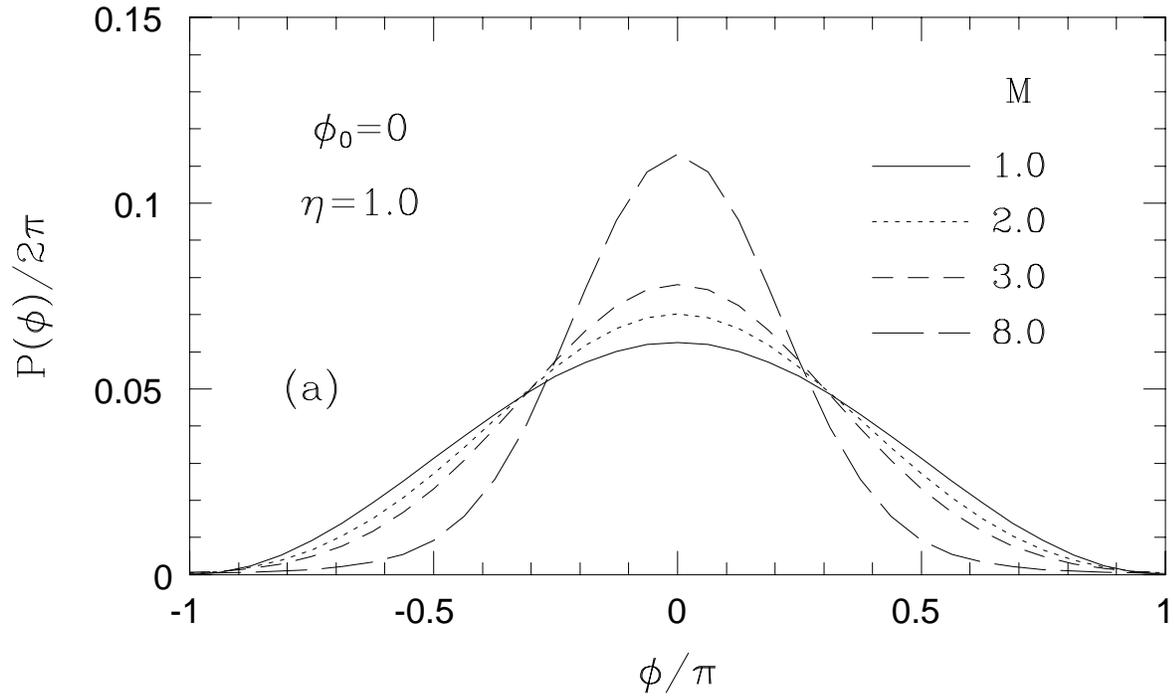
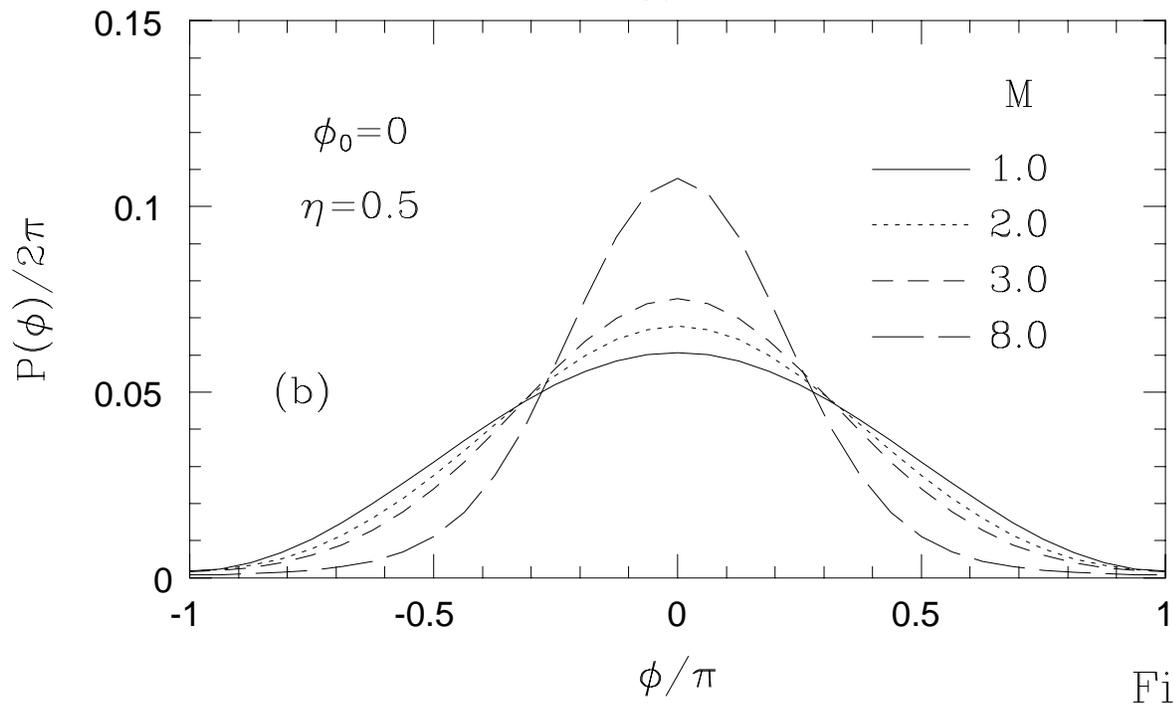
{\bf Fig.10} Measured probability distribution $P(\phi)$
versus $\phi$ for the fully polarized Fock state for 
a) $\eta=1$, b) $\eta=0.5$ and for the indicated 
for the average total photon numbers.
\\
\\
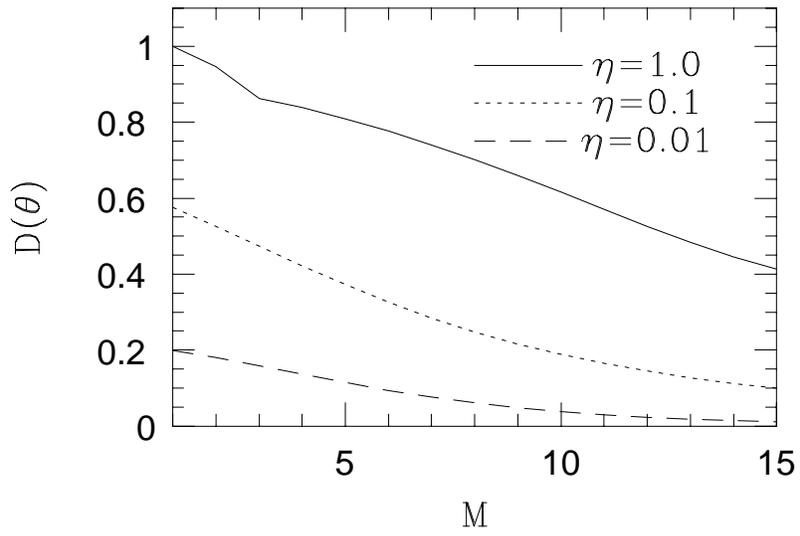
{\bf Fig.11} Second-order fluctuations in the $\theta$  
related measurements for the fully polarized Fock state 
as a function of the average total number of photons and 
for the indicated values of $\eta$ (here we considered $\phi_0=0$). 
\\
\\
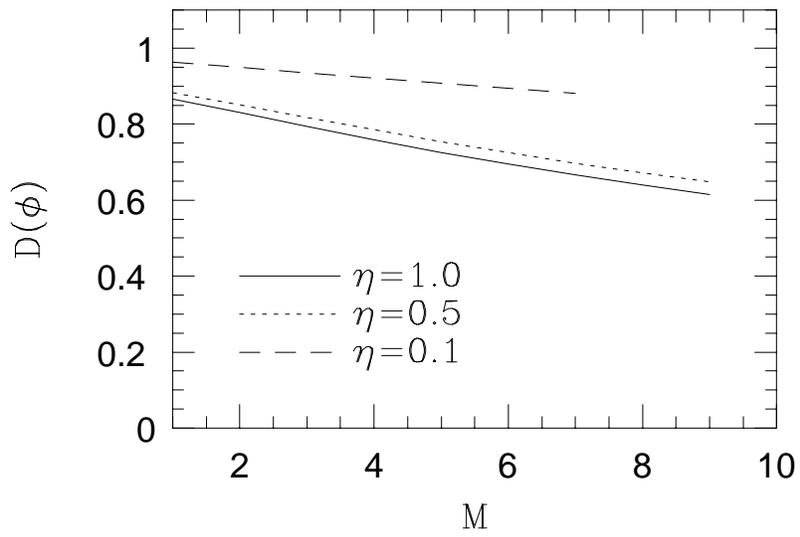
{\bf Fig.12} Second-order fluctuations in the $\phi$
related measurements for the fully polarized Fock state 
as a function of the total number of photons and 
for the indicated values of $\eta$ (here we considered $\phi_0=0$).
\\
\\
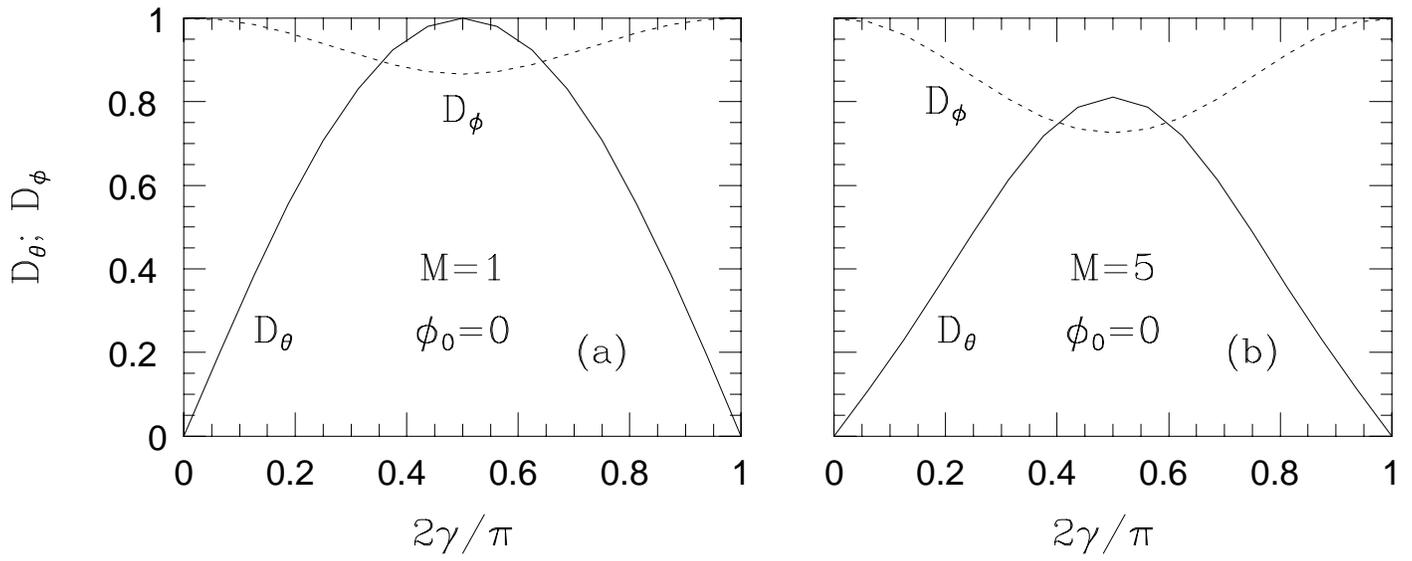
{\bf Fig.13} Second-order fluctuations in the $\theta$ and $\phi$
related measurements for the fully polarized Fock state 
as a function of the rotation parameter $\gamma$
for the indicated values of $M$ and $\phi_0$.
\\
\\
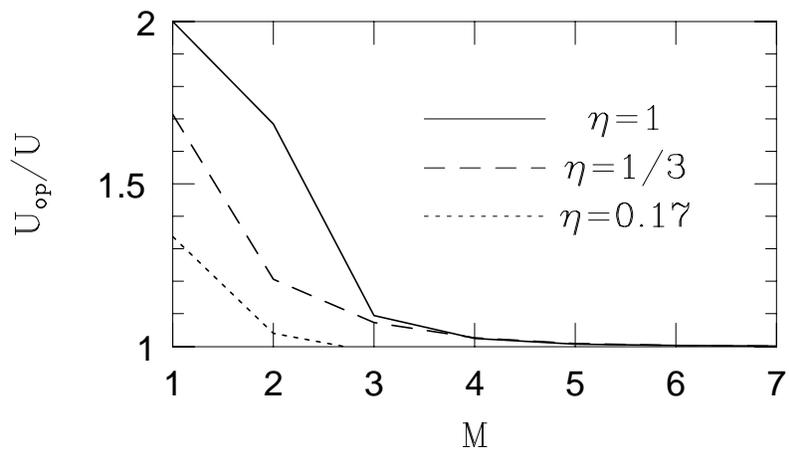
{\bf Fig.14} Comparison between the theoretical uncertainty product 
and the measured one in the fully polarized Fock state for the indicated 
values of $\eta$. 
\end{document}